\documentclass[fleqn,usenatbib]{mnras}
\usepackage{newtxtext,newtxmath,multirow}
\usepackage[T1]{fontenc}
\DeclareRobustCommand{\VAN}[3]{#2}
\let\VANthebibliography\thebibliography
\def\thebibliography{\DeclareRobustCommand{\VAN}[3]{##3}\VANthebibliography}

\usepackage{graphicx}
\usepackage{amsmath}
\usepackage{booktabs}
\usepackage{bm}
\usepackage{url}

\newcommand{\nd}{$\dots$}

\DeclareRobustCommand{\okina}{\raisebox{\dimexpr\fontcharht\font`A-\height}{\scalebox{0.8}{`}}}

\title[M33 Miras \& LPVs in \textit{griJHK$_S$}]{The M33 Synoptic Stellar Survey. III. Miras and LPVs in {\it griJHK$_S$}}

\author[Konchady et al.]
{Tarini~Konchady $^{1}$\thanks{tkonchady.tamu@gmail.com} \href{https://orcid.org/0000-0003-0452-9182}{\includegraphics[width=0.025\textwidth]{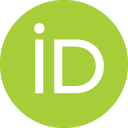}},
Lucas~M.~Macri$^{2}$\href{https://orcid.org/0000-0002-1775-4859}{\includegraphics[width=0.025\textwidth]{orcid.png}},
Xiaomeng~Yan$^{3}$\href{https://orcid.org/0000-0002-6375-6310}{\includegraphics[width=0.025\textwidth]{orcid.png}}, 
and Jianhua~Z.~Huang$^{4}$\href{https://orcid.org/0000-0002-7735-3002}{\includegraphics[width=0.025\textwidth]{orcid.png}}
\\
$^{1}$The National Academies of Sciences, Engineering, and Medicine, Washington, DC, USA\\
$^{2}$NSF National Optical-Infrared Astronomy Research Laboratory, Tucson, AZ, USA\\
$^{3}$Capital University of Economics and Business, Beijing, China\\
$^{4}$School of Data Science, Chinese University of Hong Kong, Shenzhen, China
}

\date{Accepted 2024 March 14. Received 2024 March 13; in original form 2023 September 30}
\pubyear{2024}

\begin{document}
\label{firstpage}
\pagerange{\pageref{firstpage}--\pageref{lastpage}}
\maketitle

\begin{abstract}
We present the results of a search for Miras and long-period variables (LPVs) in M33 using $griJHK_S$ archival observations from the Canada-France-Hawai\okina i Telescope. We use multiband information and machine learning techniques to identify and characterize these variables. We recover $\sim$1,300 previously-discovered Mira candidates and identify $\sim$13,000 new Miras and LPVs. We detect for the first time a clear first-overtone pulsation sequence among Mira candidates in this galaxy. We use O-rich, fundamental-mode Miras in the LMC and M33 to derive a distance modulus for the latter of $\mu=24.629\pm0.046$~mag.
\end{abstract}


\section{Introduction} \label{sec:intro}

The local measurement of the Hubble Constant (H$_0$) by \citet{riess22} differs by $>\!5\sigma$ from the value expected from observations of the Cosmic Microwave Background and Baryon Acoustic Oscillations under the assumption of $\Lambda$CDM \citep{planck}. Local measurements of H$_0$ are often based on Cepheid variables (Cepheids) and Type Ia supernovae (SNe~Ia) as primary and secondary distance indicators, respectively. Additional independent primary distance indicators can increase the number of secondary distance indicators or enable new distance ladders to better characterize this tension.

Mira variables (hereafter, Miras) can serve as one of these primary distance indicators. Miras are Asymptotic Giant Branch (AGB) stars that can pulsate in fundamental or overtone modes \citep{Wood1996} with typical periods ranging from $\sim$100 to $\sim$3,000~days \citep{ogle05,ogle09,riebel2010}. Mira variability is cyclic, characterized by large {peak-to-trough} amplitudes at optical wavelengths (historically, photo-visual magnitude amplitude $>2.5$~mag, see \citealt{kukarkin1958,clayton69}; more recently $\Delta I > 0.8$ mag, \citealt{ogle09}) and long-term variations in mean magnitude (\citealt{mattei1997}, \citealt{whitelock1997}). Miras also vary in the near-infrared (NIR) with smaller amplitudes ($\Delta K_S > 0.4$ mag, \citealt{whitelock2008}). Given their low- to intermediate-mass progenitors ($0.8 M_{\odot} < M < 8 M_{\odot}$; \citealt{whitelock2013}), they are common and can be found across all types of galaxies. The Milky Way and the Magellanic Clouds have proven to be prodigious sources of Miras, as revealed by the Optical Gravitational Lens Experiment (OGLE; \citealt{udalski1992}) and the MACHO project \citep{alcock1993}.

Miras are typically classified as Oxygen- or Carbon-rich (hereafter, O- and C-rich) based on the dominance of Oxygen- or Carbon-rich molecules in their spectra, affected by the CNO cycle, helium shell burning, and other internal stellar processes. $^{12}$C and $^{18}$O in particular can be ``dredged up" and raised to the surface of the star \citep{iben_renzini1983, kobayashi2011, karakas2014}. The occurrence of dredge-up events is dictated by stellar mass. In the case of stars with $M > 4 \ M_{\odot}$, another phenomenon called ``hot-bottom burning" (HBB) comes into play. During HBB, the bottom of the convective layer heats up to the point that the CNO cycle is activated with the rare appearance of the Na-Na and Mg-Al cycles. This can affect the transition from O-rich to C-rich for Miras, with some models allowing for C-rich Miras to be converted back to O-rich ones (\citealt{hinkle2016}, \citealt{whitelock2000}).

Miras follow tight NIR Period-Luminosity relations \citep[PLRs;][]{glass_evans1981,glass_feast_1982}. In the Large Magellanic Cloud (LMC), O-rich Miras have $K$-band PLRs with low scatter ($\sigma = 0.12$~mag; \citealt{yuan2017b}) that is comparable to the scatter of Cepheid PLRs in the same band ($\sigma = 0.09$~mag; \citealt{macri2015}). 

O-rich Miras with $P<400$~d have been demonstrated to be useful as extragalactic distance indicators. \citet{yuan2017a} used $I$-band observations from \citet{macri2001} and \cite{pellerin_macri2011} to identify 1,847 Mira candidates in M33. Their study was extended in \citet{yuan2018} with sparsely-sampled \textit{JHK$_{S}$} light curves, where they obtained NIR PLRs for O-rich Miras and a distance modulus of $24.80\pm0.06$~mag for M33.

\citet{chuang2018} used NIR \textit{Hubble Space Telescope} (\textit{HST}) observations to identify a sample of 139 O-rich Mira candidates in NGC$\,$4258, which they coupled with LMC Miras to obtain a relative distance modulus that was consistent with Cepheid-based measurements. \citet{chuang2020} used NIR \textit{HST} observations to identify 115 O-rich Mira candidates in NGC$\,$1559 and determine its distance, in conjunction with the maser distance to NGC$\,$4258 and its sample of Miras. \citet{chuang2020} also presented a Mira-based determination of H$_0$ within $1\sigma$ of the contemporaneous Cepheid-based value from \citet{riess2019}.
\clearpage
The Vera C.~Rubin Observatory will soon begin its Legacy Survey of Time and Space (LSST), a decade-long deep time domain survey of $\sim$20,000~sq.~deg.~in the \textit{ugrizY} bands \citep{lsst}. \citet{wenlong_thesis} transformed LMC Mira PLRs (from $VI$ to $griz$) to estimate that LSST should yield $\sim$200,000 Miras across $\sim$200 galaxies within $\sim$15 Mpc, with $\sim$75 of these yielding upwards of 100 Miras each.

In light of this, detailed characterization of Mira properties in \textit{griz} would benefit searches for Miras in LSST. \citet{ou2023} used multiple \textit{gri} surveys with a baseline of $\sim$18 years to improve the periods of 1,637 previously-discovered Miras in M33. They used transformed O-rich Mira $i$-band magnitudes at maximum light to derive a distance modulus of $24.67\pm0.06$~mag for M33. They also noted that in order to accurately determine Mira periods, it is vital to obtain full-amplitude light curves, as opposed to relying on samples around maximum light.

 In this paper, we present the results of a Mira search using $griz$\textit{JHK$_{S}$} observations of M33 and informed by the results presented in \citet{yuan2017a} and \citet{yuan2018}. 

 $\S$\ref{sec:obs_data_red} describes our observations and photometry, $\S$\ref{sec:mira_id} lays out our procedure for identifying Mira candidates using NIR information, and $\S$\ref{sec:id_new_miras} describes our attempts to use machine learning methods and LMC long-period variables (LPVs) to identify new Mira candidates.
\vspace{-18pt}
\section{Observations and Data Reduction}\label{sec:obs_data_red}

\subsection{MegaCam and WIRCam Observations}\label{megacam_obs}
We used archival pipeline-processed optical observations of M33 taken with the MegaCam instrument \citep{megacam} on the Canada-France-Hawai\okina i Telescope (CFHT). The observations were acquired as part of proposal IDs 04BF26 (PI Beaulieu) and 04BH98 (PI Hodapp). Results from the former were originally presented by \citet{hartman2006}. The data were obtained using the $gS$, $rS$, $iS$, and $zS$ filters\footnote{\href{https://www.cfht.hawaii.edu/Instruments/Imaging/Megacam/specsinformation.html}{www.cfht.hawaii.edu/Instruments/Imaging/Megacam/\\specsinformation.html}} (hereafter \textit{griz}) with a baseline of roughly one-and-a-half years (August 2003 to January 2005).

MegaCam is a wide-field (1 deg.~on a side) optical imager consisting of 36 CCDs with a plate scale of $0\farcs187$ per pixel. Each frame is a mosaic image as a result of the CCD array (see Fig.~\ref{fig:m33_ccds}). We split each frame into one image per individual CCD and then sorted the images by band. We visually inspected each image and discarded any unusable ones (e.g., due to poor image quality). This yielded a typical coverage of 29, 27, 28 and 1 nights in $griz$, respectively.

We also used archival pipeline-processed NIR observations of M33 obtained with the Wide-field InfraRed Camera (WIRCam; \citealt{wircam}) on CFHT. The observations were taken as part of three different studies (proposal IDs 06BF36 \& 07BF23, PI Beaulieu, $JK_S$, 2006-07; proposal ID 15BT03, PI Ngeow, $H$, 2015; proposal ID 17BT02, PI Lee, $H$, 2017-18) and covered different areas within the central disk of M33. The approximate spans of the programs were 1, 2.5 and 1 year, respectively.

WIRCam consists of four detectors with a combined field of view $20\farcm5$ on a side and a plate scale of $0\farcs3$ per pixel (\citealt{wircam}). Each WIRCam frame is effectively a ``data cube" comprised of four to five 10-second exposures of the field of view at the time of observation. Each exposure is a mosaic of four images corresponding to each of the chips. We combined the multiple 10-second exposures of a given chip within each frame and only carried out photometry on these composite images. This yielded an average of 9, 6 and 3 epochs in \textit{JHK$_S$}, respectively, for locations imaged in a given band.

The cadence of the MegaCam and WIRCam observations are shown in Fig.~\ref{fig:cadence}, while Fig.~\ref{fig:source_avail} shows the cumulative distribution of detected sources as a function of the number of the epochs available for that band. As we will show later, our typical Mira candidates had 4, 13, 43, 1, 6, 5 and 2 observations in $griz$\textit{JHK$_{S}$}, respectively.

\subsection{Photometry}
To process the MegaCam images, we first identified a reference epoch for each band by examining the point-spread functions (PSFs) of stars in the images associated with CCD\#11. We chose that detector because it partially covers the disk of M33 at a reasonable source density. None of the reference epochs had any unusable images.

We obtained aperture and PSF photometry for all images using {\fontfamily{pcr}\selectfont DAOPHOT}, {\fontfamily{pcr}\selectfont ALLSTAR}, {\fontfamily{pcr}\selectfont ALLFRAME} and related programs \citep{stetson1987, stetson1994} with a Python wrapper. A primary image for each CCD and band was constructed using {\fontfamily{pcr}\selectfont MONTAGE}. The primary images were then used to create source lists for {\fontfamily{pcr}\selectfont ALLFRAME}. {\fontfamily{pcr}\selectfont TRIAL} \citep{stetson1996} was used to perform frame-to-frame zeropoint corrections, calculate variability statistics, obtain mean instrumental magnitudes, and extract light curves. Sources were then matched across filters for each CCD. The photometric uncertainties versus magnitude for each band are shown in Fig.~\ref{fig:phot_uncer}.

\begin{figure*}
 \includegraphics[width=0.87\textwidth]{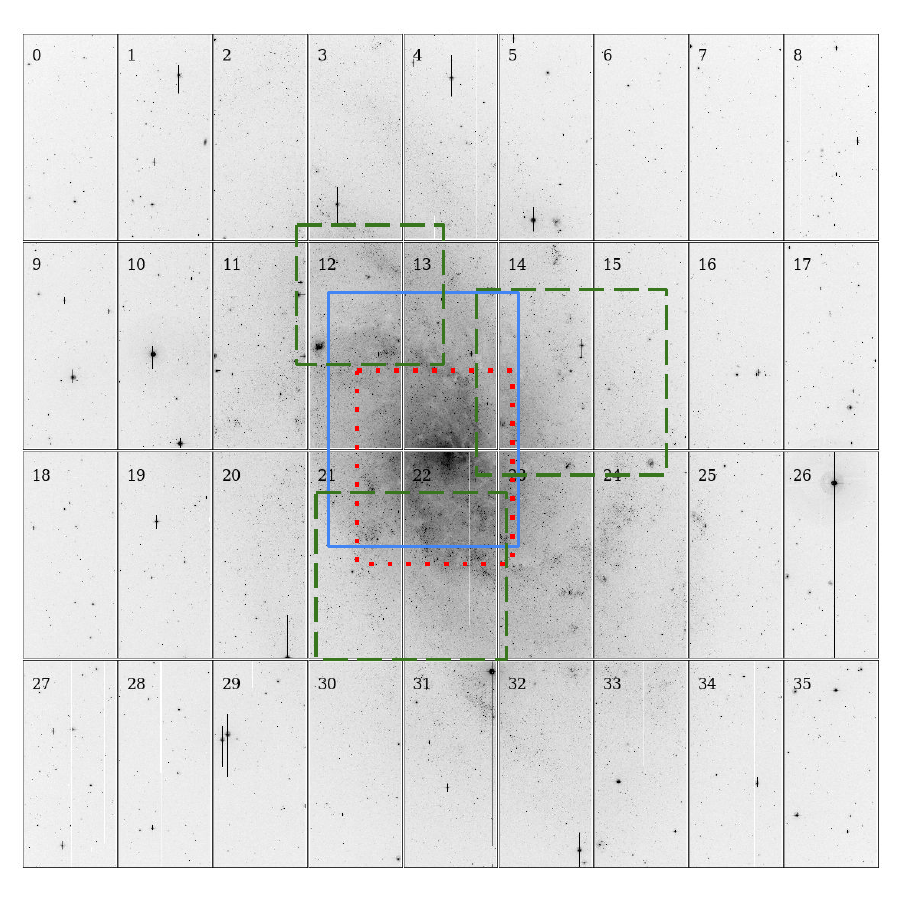}
 \caption{Mosaic of a typical CFHT MegaCam image of M33 with the CCD numbers marked, also showing the areas covered by WIRCam in $J$ (blue, solid), $H$ (green, dashed), and $K_S$ (red, dotted) fields. North is up and east is to the left.}
 \label{fig:m33_ccds}
\end{figure*}

We performed the astrometric and photometric calibration of the MegaCam sources using images from the Panoramic Survey Telescope and Rapid Response System (Pan-STARRS) Telescope \#1 Data Release 1 (PS1 DR1; \citealt{panstarrs}). We solved for the astrometric solution using {\fontfamily{pcr}\selectfont WCSTools} \citep{mink} with the primary image of each CCD and filter. Following this, we matched CFHT and PS1 DR1 sources with a tolerance of 2\arcsec. If multiple sources satisfied that criterion, the closest Pan-STARRS source was selected.

We used our list of astrometrically-calibrated sources to solve for the following photometric transformations with iterative 2.5$\sigma$ rejection:
\begin{equation}
\label{griz_phot_calibration_eqn}
 m_{C} - m_{I} \ = \textrm{ZP} + \chi + \xi (\textrm{col} - \text{piv})
\end{equation}
\noindent where $m_{C}$ is the fully-calibrated PS1 magnitude, $m_{I}$ is the instrumental magnitude reported by {\fontfamily{pcr}\selectfont DAOPHOT/ALLSTAR/ALLFRAME} (corrected for exposure time), ZP is the MegaCam default zeropoint for a given band\footnote{\href{https://www.cfht.hawaii.edu/Instruments/Imaging/MegaPrime/generalinformation.html}{www.cfht.hawaii.edu/Instruments/Imaging/MegaPrime/\\generalinformation.html}}, $\chi$ is the residual zeropoint, $\xi$ is the color term, col is the PS1 color, and piv is a ``pivot'' color value typical of our target stars. We solved for global values of $\xi$ for each transformation, using several thousand stars spanning a wide range of colors, and for chip-specific values of $\chi$ using $\sim$150 stars per CCD, with a typical scatter of 0.04~mag.

To process the WIRCam images, we first identified the fields associated with the various observing programs. There was no consistent overlap across all the frames and filters, as seen in Fig.~\ref{fig:m33_ccds}. Before beginning photometry, we separated the images into groups based on their location on the sky (4, 25 and 10 groups for $JHK_S$, respectively). Reference images for each group were chosen by visual inspection. We then obtained aperture and PSF photometry for the WIRCam images using the same methods as for the MegaCam images.

We performed the astrometric and photometric calibration of the WIRCam sources using the catalog from \citet{javadi2015}, based on observations with the UK InfraRed Telescope (UKIRT). We used the WIRCam WCS information for a preliminary match against the UKIRT catalog, finding global offsets of $0.5-1.5\arcsec$ (depending on the band) and correlated residuals as a function of position. These are likely due to different geometrical distortions in the two cameras and were removed using second or third-order polynomials. The remaining residuals were removed using a non-parametric technique. We divided each image in $100\times100$ pixel cells and used the average residuals of the stars in each cell to fit a thin-plate spline and apply the necessary correction. The final dispersion of position residuals after all corrections was $\sim0\farcs1$. All corrections done after the initial global offset were based on the 5,000 brightest stars in a given image.

\ \par

We solved for the the NIR photometric transformations using the following equation with iterative $2.5\sigma$ rejection:

\begin{equation}
\label{nir_phot_calibration_eqn}
 m_{I} - m_{C} \ = \chi \ + \xi (J-K_S - 1.0) \ + \xi^{'} (J-K_S - 1.0)^2 
\end{equation}
\ \par
\ \par
\ \par

\noindent where $m_{I}$ is the instrumental magnitude, $m_{C}$ is the calibrated UKIRT magnitude, $\chi$ is the residual zeropoint\footnote{\href{www.cfht.hawaii.edu/Instruments/Imaging/WIRCam/WIRCamThroughput.html}{www.cfht.hawaii.edu/Instruments/Imaging/WIRCam/\\WIRCamThroughput.html}}, $\xi$ and $\xi^{'}$ are the first- and second-order color terms and $J-K_S$ is the UKIRT color. We solved for global parameters across the four detectors.

The mean values of all coefficients in the photometric transformations are presented in Appendix Tables~\ref{tab:pan_trans_coeff} and \ref{tab:nir_trans_coeff}, and representative solutions are shown in Fig.~\ref{fig:trans_resids}. These were applied to all stars in our catalog, including those lying outside of the color range spanned by the local standards. We fully propagated the uncertainties in all transformation coefficients, which for our objects of interest never exceeded 0.03~mag.

Appendix Table~\ref{tab:calphot} presents the fully-calibrated time-averaged magnitudes in all available bands for all $\sim$1.15 million point sources detected in our analysis, as well as their $gri$ variability statistics.

\clearpage

\begin{figure*}[t]
\begin{center}
 \includegraphics[width=0.76\textwidth]{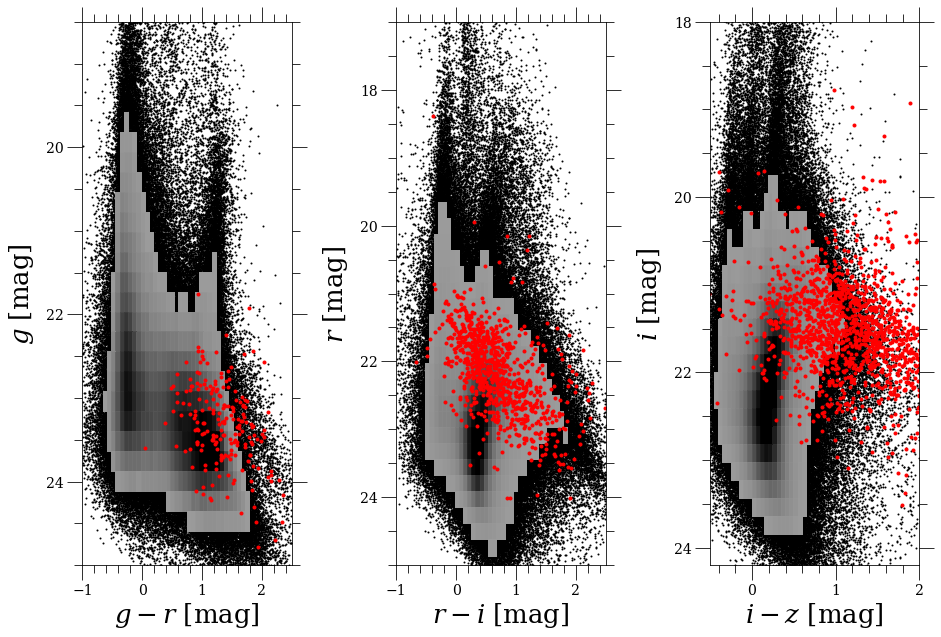}
\end{center}
\caption{Hess diagrams in the optical bands. Individual stars are plotted where the source density drops below 200 objects per bin. Recovered Miras from \citet{yuan2017a} are shown using red points. Mira recovery varies across filters. Not all recovered Miras were kept in our final samples, due to quality cuts.}
\label{fig:hess_optical}
\end{figure*}

\begin{figure*}[b]
\begin{center}
\includegraphics[width=0.8\textwidth]{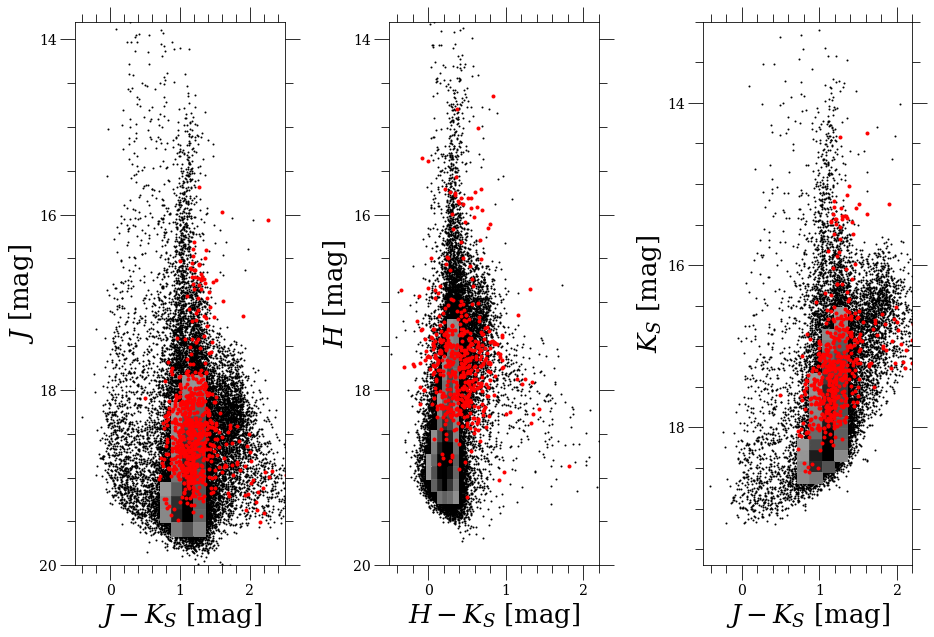}
\end{center}
\caption{Same as Fig.~\ref{fig:hess_optical}, but for the NIR bands.}
\label{fig:hess_nir}
\end{figure*}

\clearpage

\subsection{Crowding corrections}

Biased magnitude measurements have long been recognized as an issue affecting crowded-field photometry of faint stars \citep{mcclure1985,stetson1987,schechter1993}. The standard approach to characterizing and correcting this bias relies on the injection of artificial stars in the vicinity of, and with the same flux as, the objects of interest. Only a few artificial stars should be added around a given object, in order to mitigate any further crowding of the image.

\vspace{6pt}

Following these precepts, we created 20 copies of each $JHK_S$ primary image and in each one we added 5 artificial stars to the vicinity of each of the 14,312 variables described in \S3.

\vspace{6pt}

The radial distances between artificial stars and their corresponding variables were drawn from a uniform random distribution spanning 5-12~pix (1.5-3.6\arcsec). Every artificial star in a given copy of a primary image had to be at least 4 pix (1.2\arcsec) away from each other and from every Mira candidate or other stars in the frame (down to 2~mag fainter than the corresponding source of interest).

\vspace{6pt}

We carried out the same photometric procedures on each of the artificial images as previously done in the real ones, identified the artificial sources and compared the recovered and input magnitudes. The results are shown in Fig.~\ref{fig:crowd}. Crowding corrections are minimal for objects with $m<$19: $0.014\pm0.054$ ($J$), $0.009\pm0.067$ ($H$), $-0.002\pm0.042$ ($K_S$). $\sim 22$\% of the variables had crowding corrections exceeding 0.1~mag or uncertainties in those corrections beyond 0.1~mag; they were flagged accordingly and excluded from the final samples.
\section{Identifying LPVs and Mira Candidates Using Optical and NIR Observations}\label{sec:mira_id}

\subsection{Initial selection}

We identified $\sim$1.15 million unique objects from our photometry which had at least one detection in one of the $gri$ bands. The Hess/color-magnitude diagrams for the optical and NIR bands are shown in Figs.~ \ref{fig:hess_optical} and \ref{fig:hess_nir} respectively, with the recovered Miras from \citet{yuan2017a} and \citet{yuan2018} overplotted.

\vspace{6pt}

We began the selection of LPVs and Mira candidates by making variability, color, and amplitude cuts based on the optical data. The variability cuts were based on the Stetson $J$ index \citep{stetson1996} calculated from our $i$ band measurements ($J_i$) using {\fontfamily{pcr}\selectfont TRIAL}. This index takes into account correlated deviations from a mean magnitude and their measurement quality; the higher its value, the more likely an object is genuinely variable. We only considered objects with $J_i \geq 0.75$, which corresponds to a $\sim$5$\sigma$ detection of variability using this index. Fig.~\ref{fig:ji_hist} presents a histogram of $J_i$ values while Fig.~\ref{fig:ji_vs_i} shows $J_i$ versus $i$. As the next step, since Miras are red variables, we only considered objects with either $r-i \geq 0$ or a non-detection in the $r$ band. Among the remaining objects, we selected those whose $i$-band light curves spanned a range ($R_i$) of at least 0.3~mag (see Fig.~\ref{fig:ai_hist} for the overall distribution of this parameter). We selected this threshold as it only excluded $\sim$1\% of the previously-known Mira candidates recovered by our photometry.

\vspace{6pt}

We then selected the objects that were detected in at least three epochs in one of the NIR bands, as that information is needed for our subsequent analysis. Following all the stated cuts (summarized in Table~\ref{tab:cuts}), we were left with 14,312 variables, which included 1,342 of the Miras identified in \citet{yuan2018}.

\subsection{Light curve fits}

We fit the available {\it griJHK}$_S$ light curves of these variables using a simple sinusoidal model, defined for a given band as
\begin{equation}
 \label{light_curve_fit_eqn}
 m(t_i) \ = \ \overline{m} \ - \ A \sin( {2\pi}t_i/{P}+ \phi) 
\end{equation}
where $m$ is the magnitude at time $t_i$, $\overline{m}$ is the mean magnitude, $A$ is the semi-amplitude, $P$ is the period, and $\phi$ is a phase offset. Since the $z$ measurements were obtained on a single night, they were not considered further in the analysis. We simultaneously fit light curves from several bands using MPFIT \citep{markwardt2009}; the values of $\overline{m}$, $A$ and $\phi$ might be different for each band but $P$ was always solved for as a common parameter. We fit each variable using 55 trial periods equally spaced every $0.02 \log P$, spanning $1.925\leq\log P \leq 3.005$, and selected the fit with the lowest $\chi^2_\nu$.

We attempted to fit the light curves using modern techniques, such as the semi-parametric Gaussian Process model \citep{yuan2017a, he2016} and stochastic variational inference models \citep{he2021} that have been used recently on longer Mira time series. Unfortunately, the limited number of cycles covered by our data and the lack of time overlap between the optical and the NIR bands hampered the performance of these models.

We initially fit only the $iJHK_S$ light curves for the 1,342 recovered Miras from \citet{yuan2017a}, solving for independent values of $\overline{m}$, $A$ and $\phi$. Using objects from this subsample with information in at least two NIR bands, we found $A_H \sim$ $A_J$, $A_{K_S} \sim$ $0.9 A_J$, $\phi_H \sim$ $\phi_J$ and $\phi_{Ks} \sim$ $\phi_J - 0.03$, which we adopted for the subsequent analysis. This simplification was motivated by the relatively small number of data points per light curve in the NIR bands.

Having implemented the above interrelations for NIR amplitudes and phases, we next expanded the fit to all six bands for objects in this subsample. We further derived $\phi_r \sim \phi_i + 0.015$ and $\phi_g \sim \phi_i + 0.03$, which we also adopted for the subsequent analysis. We did not find useful interrelations for the amplitudes in $gri$, nor could we find a robust global offset between $\phi_i$ and $\phi_J$.

Fig.~\ref{fig:compare_y17a_periods} compares the periods of objects in this subsample derived by \citet{yuan2017a} and by our procedure. We found good agreement (defined as $\Delta P < 50$~d) for $\sim$78.1\% of the objects{, with most of the others lying along the $\pm 1/365$~d alias relations. This recovery rate is very similar to the one found by \citet{yuan2018} based on simulations of Mira light curves with similar sampling to our work.} Since our fitting procedure did not return $\sigma(P)$ for all objects, we used the scatter about this 1:1 relation to determine $\langle\sigma({\Delta P}/{P})\rangle=0.03$. We only used our best-fit periods in all subsequent analysis.

\begin{figure*}
\begin{center}
\includegraphics[width=\textwidth]{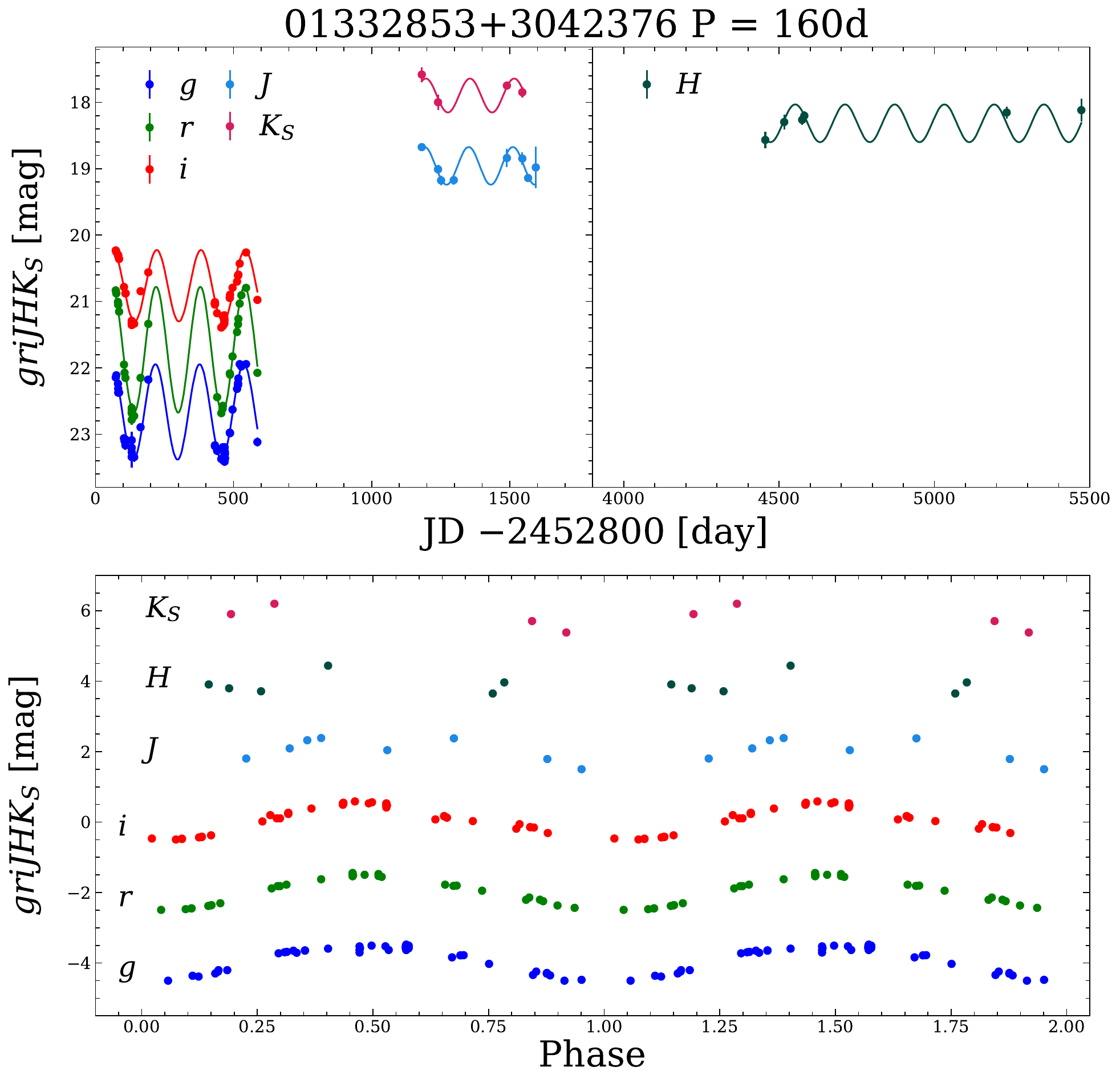}
\end{center}
\caption{A representative Mira from our sample that was previously identified by \citet{yuan2017a}. Upper panel: observed light curves in $griJHK_S$. Lower panel: phased light curves; magnitudes have been offset and two cycles are plotted for clarity.}
\label{fig:eg_mira_lcs}
\end{figure*}

We fit the coupled sinusoidal model to all 14,312 variables (using the same set of trial periods described above) to derive $P$ for each object and the following properties (when available): up to six mean magnitudes ($griJHK_S$), four amplitudes ($A_g, A_r, A_i$ and $A_{JH}$, with $A_{K_S}$ coupled to the latter), and two phases ($\phi_i$ and $\phi_{JH}$, with $\phi_g$ and $\phi_r$ coupled to the former and $\phi_{Ks}$ coupled to the latter). Fig.~\ref{fig:eg_mira_lcs} shows the light curves and best-fit model for a recovered Mira. 

Once mean magnitudes had been obtained, we calculated NIR Wesenheit indices \citep{madore1982}, defined as
\begin{equation}
W_{JK_S}=K_S - R_{JK_S}^K (J - K_S);\ \ \ W_{JH}=H - R_{JH}^H (J - H),\label{eq:wes}
\end{equation}
which simultaneously minimize the effects of temperature and extinction. We adopted $R_{JK}^K = 0.742$ and $R_{JH}^H = 1.727$, calculated as in \citet{yuan2018} using \citet{schlafly} values.

Appendix Table~\ref{tab:lpvprop} presents the properties and associated uncertainties of all 14,312 LPVs, as well as the machine-learning classification scores and other labels to be described in \S4.

\clearpage

\clearpage


\begin{figure}
\includegraphics[width=0.45\textwidth]{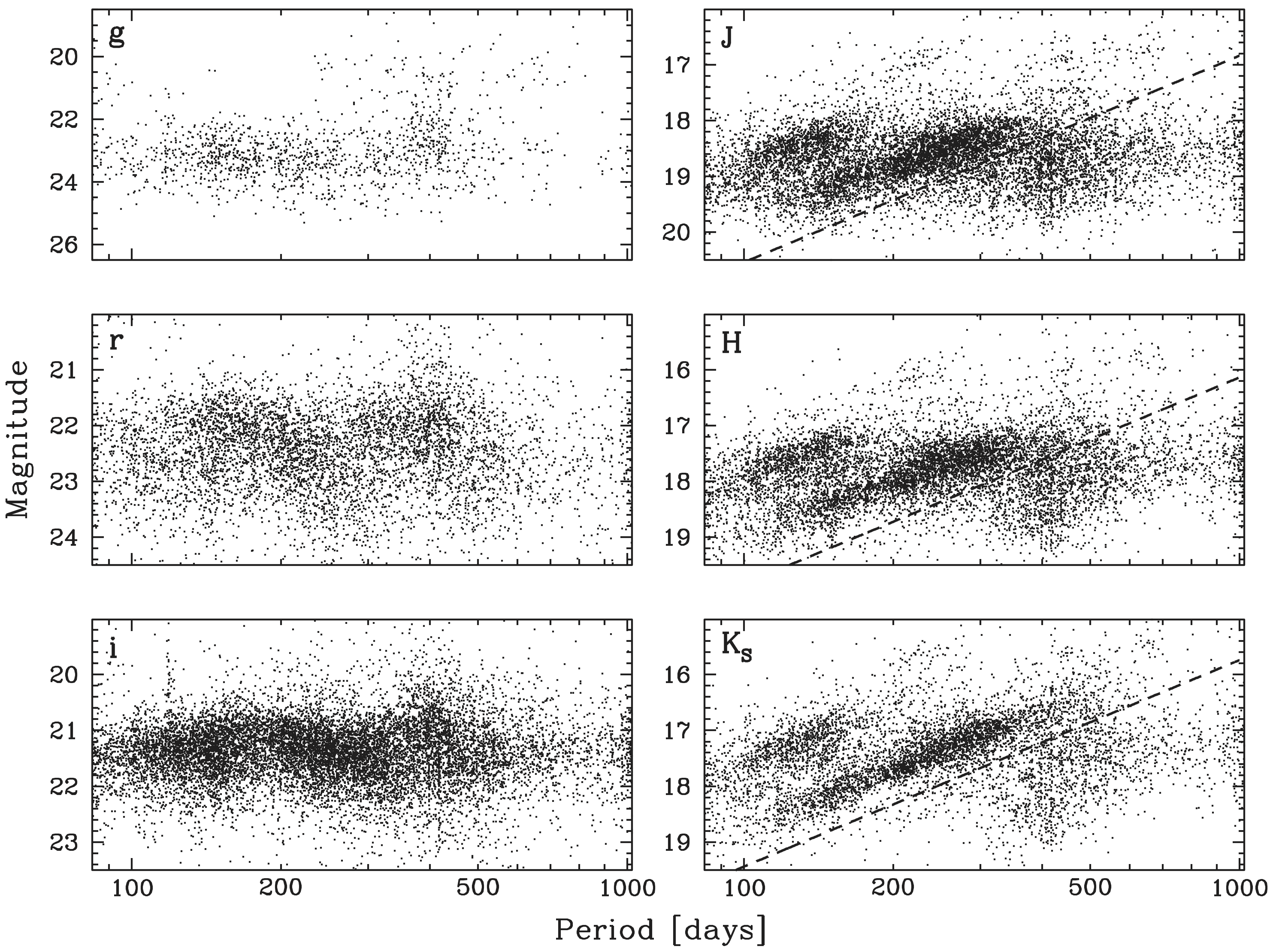}
\centering
\caption{Period-magnitude diagrams of LPVs in M33 identified in our analysis. Left column, top to bottom: $g$, $r$, $i$. Right column, top to bottom: $J$, $H$, $K_S$. The dashed lines on the right panels indicate the limits used to separate faint objects from first-overtone and fundamental pulsators.}
\label{fig:mirc_PL_diagram}
\end{figure}

\begin{figure}
\includegraphics[width=0.45\textwidth]{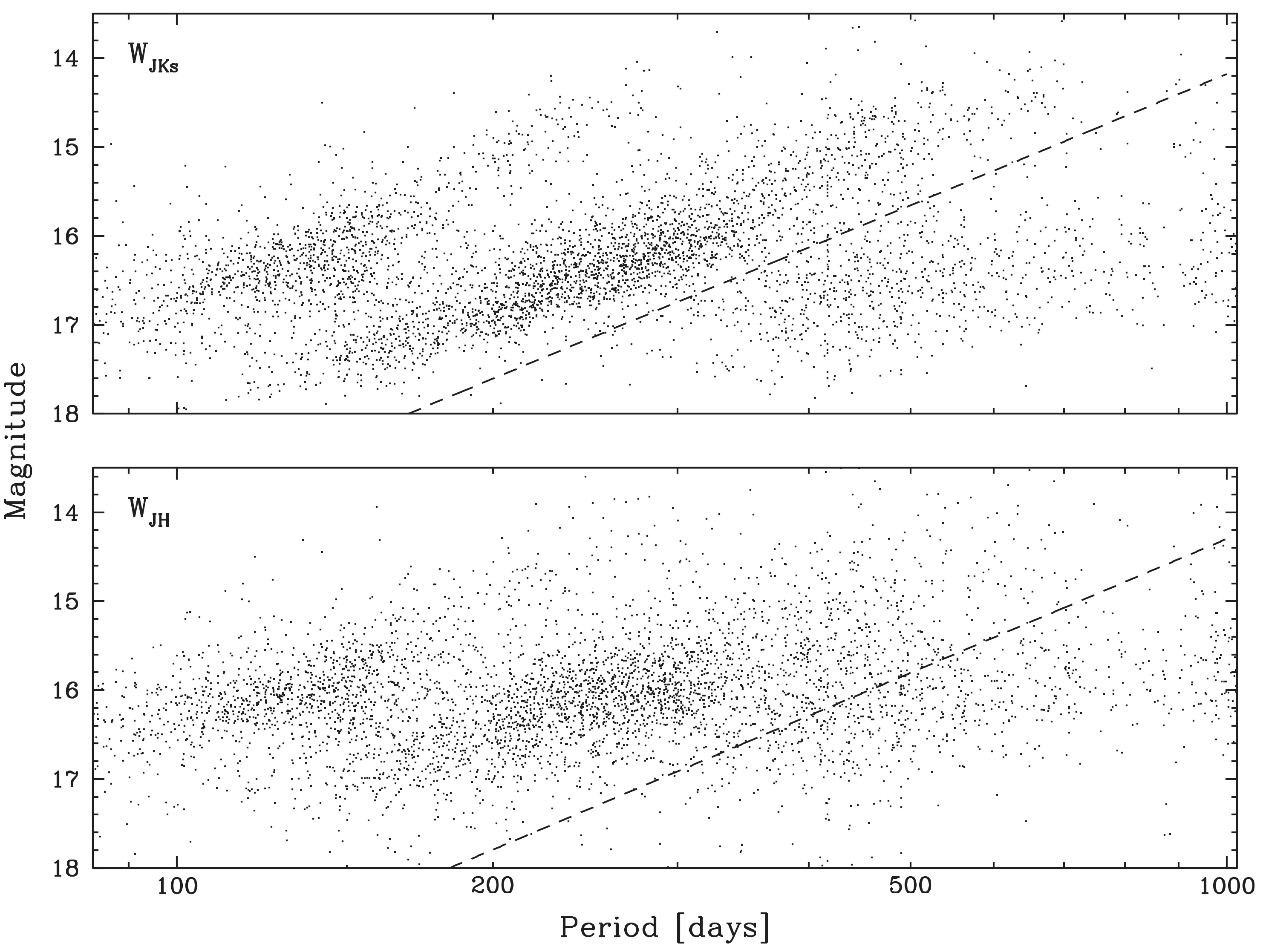}
\centering
\caption{Same as Fig.~\ref{fig:mirc_PL_diagram}, but using NIR Wesenheit indices.}
\label{fig:mirc_PW_diagram}
\end{figure}

Figs.~\ref{fig:mirc_PL_diagram} and \ref{fig:mirc_PW_diagram} show the resulting period-magnitude and period-Wesenheit diagrams. Two obvious sequences can be seen in all the NIR relations, with $K_S$ magnitudes of $\sim$ 16.8 and 18.1 at $\log P=2.2$. These correspond to first-overtone (FO) and fundamental-mode (FU) pulsators, respectively, as first identified in the LMC by \citet{Wood1996}. A third much more diffuse group (which we labeled FA for ``faint") can be seen at $\log P\gtrsim$2.6 and $K_S\gtrsim$17, becoming somewhat more obvious in the Wesenheit relations. Most of these objects have smaller values of $A_i$ compared to those in the other sequences, and may be the counterparts of the LMC LPVs plotted in green and lying below (fainter than) sequence D in Fig.~1 of \citet{Soszynski2013}.

\subsection{Classification into O- and C-rich subtypes\label{sec:colcol}}

As previously mentioned, O-rich Miras have been shown to obey tighter Period-Luminosity relations than their C-rich counterparts. Thus, it is of interest to classify Miras into subtypes in order to obtain better distance estimates. In the absence of spectroscopic observations, which would be prohibitive to obtain for large extragalactic samples, this classification must rely on photometric information.

The LMC is an ideal system to derive photometric-based classification methods for Miras given the better quality light curves that can be obtained for its variables. \citet{ogle05} combined their OGLE-II/III $V$ and $I$ photometry of LPVs with 2MASS PSC \citet{cutri03} $JHK_S$ magnitudes to derive Period-Luminosity relations of Miras and semi-regular variables (SRVs) in this system. They showed that these two types of pulsators follow the C and C$'$ sequences, respectively, originally identified by \citet{Wood1996} and later characterized by \citet{ita2004}. 

\citet{ogle09} presented a catalog of OGLE-III LMC LPVs based on optical observations that consists of 1,667 Miras and 11,128 SRVs. They combined their measurements with 2MASS NIR magnitudes to show that Miras can be reliably separated into O- and C-rich types in the $V-I$ vs.~$J-K_S$ plane, or by comparing optical and near-infrared Wesenheit indices. \citet{yuan2018} also showed that O- and C-rich Miras can be separated well in the $J-H$ vs.~$H-K_S$ plane. \citet{menzies19} studied AGB variables in NGC$\,$3109 and showed that a simple cut in $J-K_S$ is not a reliable method to separate O- and C-rich Miras due to changes in color as a function of abundance.

\begin{figure}
\begin{center}
\includegraphics[width=0.44\textwidth]{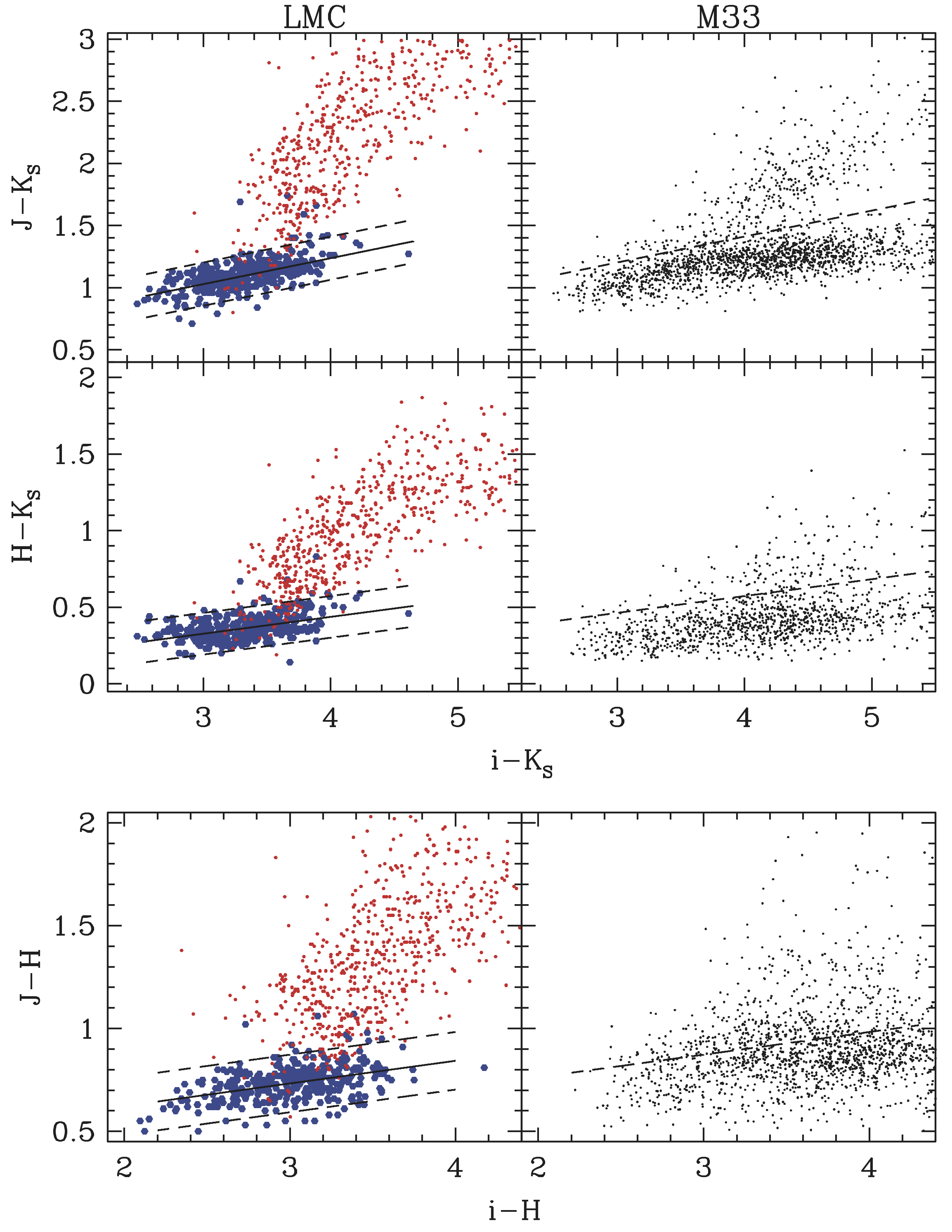}
\end{center}
\caption{Left: Color-color relations of OGLE-III Miras in the LMC. Blue and red red symbols indicate O- and C-rich variables, respectively. The solid lines are the best-fit linear relations of O-rich objects, with dashed lines indicating the $\pm2\sigma$ dispersion. Right: Same relations for our final M33 samples, with dashed lines indicating the division between O- and C-rich variables.}
\label{fig:colcolsel}
\end{figure}

\begin{table}
\centering
\begin{tabular}{llr}
\toprule
\multicolumn{1}{c}{Color}& \multicolumn{1}{c}{Relation} & \multicolumn{1}{c}{$\sigma$}\\
\midrule
$J-K_S$ & 1.028\ +\ 0.209 $(i-K_S-3)$ & 0.087\\  
$H-K_S$ & 0.327\ +\ 0.110 $(i-K_S-3)$ & 0.068\\  
$J-H$   & 0.733\ +\ 0.110 $(i-H-3)$   & 0.070\\  
\bottomrule
\end{tabular}
\caption{Color-color relations of LMC O-rich Miras, used to classify M33 variables.}
\label{tab:colcol}
\end{table}

\clearpage

Regrettably, given the nature of the archival observations of M33 that form the basis of our study, we do not have $I$ magnitudes (but have $i$ instead), only a very small fraction of Miras have $g$ magnitudes, and there is little overlap between the $J/K_S$ and $H$ observations. Fortunately, we have $iJK_S$ magnitudes for the vast majority of objects.

\vspace{9pt}

We cross-matched the OGLE-III LMC Miras with the near-infrared catalog of \citet{kato2007}, which is significantly deeper and has considerably better angular resolution than 2MASS, to obtain improved $JHK_{S}$ magnitudes. We used the BaSTI stellar evolution database \citep{basti13,basti21} to derive cubic relations between $I-JHK_S$ and $i-JHK_S$ colors for RGB and AGB stars, so that we could directly compare the LMC and M33 relations. We used the existing OGLE classification of these Miras into O- and C-rich types to generate the diagrams shown on the left-hand side of Fig.~\ref{fig:colcolsel}. The O-rich Miras delineate tight sequences in the various color-color relations, which are listed in Table~\ref{tab:colcol}, with a scatter of $\sim0.07$~mag.

\vspace{9pt}

We used the $+2\sigma$ ridge of the LMC $(i-K_S, J-K_S)$ color-color relation (top-left panel of Fig.~\ref{fig:colcolsel}) delineated by the O-rich Miras as our primary method to classify the M33 variables. If $J$-band photometry was not available, we used the equivalent ridge in the LMC $(i-K_S, H-K_S)$ relation (middle-left panel of the same Figure). If $K_S$ band photometry was not available, we used the $+3\sigma$ ridge of the LMC $(i-H, J-H)$ relation (bottom-left panel of the same figure) to account for additional dispersion in the M33 sample. If only one NIR band was available, no classification was done.

\section{Identifying New Mira Candidates}\label{sec:id_new_miras}

We identified new Mira candidates out of our LPV sample using two approaches. The first one relies on machine-learning techniques applied to our $i$ light curves, while the other one uses near-infrared Period-Luminosity relations of Miras in the Large Magellanic Cloud. We then derived period-Wesenheit relations for the newly-selected Mira candidates and the previously-known ones.

\vspace{9pt}

We applied several quality cuts to the 12,970 LPVs not previously classified as Miras before carrying out these procedures. The cuts were applied consecutively and resulted in the rejection of the following numbers of variables:
\begin{enumerate}
 \item 2,891 with crowding corrections or uncertainties in these corrections exceeding 0.1~mag;
 \item 1,336 with best-fit periods near the lower or upper limits of our grid search ($\log P<$2 or $\log P>$3);
 \item 296 with abnormally blue colors (any of $J\!-\!K_S<$0.8, $H\!-\!K_S<$0.2, $J\!-\!H<$0.5, $i\!-\!H<$2, $i\!-\!K_S<$2.5);
 \item 2,303 faint variables lying below the first-overtone and fundamental mode groups, identified as follows. The dividing lines are also plotted in Figs.~\ref{fig:mirc_PL_diagram} and ~\ref{fig:mirc_PW_diagram}.

 \begin{itemize}
 \item 783 with $W_{JK_S}>18.0\!-\!4.9\ (\log P\!-\!2.2)$;
 \item 332 with $W_{JH}>18.0-4.9\ (\log P\!-\!2.2)$;
 \item 262 with no $J$ or $H$ data and $K_S>18.6\!-\!3.7\ (\log P\!-\!2.2)$;
 \item 532 with no $J$ or $K_S$ data and $H>19.1\!-\!3.7\ (\log P\!-\!2.2)$;
 \item 394 with no $H$ or $K_S$ data and $J>19.8\!-\!3.7\ (\log P\!-\!2.2)$
 \end{itemize}

\end{enumerate}
which resulted in a classification sample of 6,144 variables. These cuts were also applied to the recovered Miras from \citet{yuan2018}; 934 out of 1,342 were selected for further analysis.

\subsection{Machine Learning Classification}\label{sec:ml_classifiers}

We used six machine learning methods as classifiers to identify new Mira candidates: logistic regression, random forest, linear discriminant analysis, quadratic discriminant analysis, kernel support vector machine (SVM), and positive-unlabeled learning with bagging SVM \citep{mordelet2014}. All six of the methods work as binary classifiers, which is ideal for our goal of distinguishing Miras from other types of variables. We set up the classifiers so that each one returned a score for each object; the higher the value, the more Mira-like.

\begin{table}
\centering
\begin{tabular}{llcr}
    \toprule
    \multicolumn{1}{c}{Feature} & \multicolumn{1}{c}{Description} & \multicolumn{1}{c}{Source} & \multicolumn{1}{c}{Rank} \\
    \midrule
    $\sigma(R_q)/$ & Ratio of std deviations, \dotfill & L  & 2 \\
    $ \sigma(\overline{m})$&defined below & \\[3pt]
    $R_{0.9}$ & Light curve range from \dotfill & L & 3 \\
    &10th to 90th percentile &\\[3pt]
    $R$ & Light curve range \dotfill & L & 5 \\[3pt]
    $A_P$ & Semi-amplitude of periodic component \dotfill & M & 6 \\[3pt]
    $\sigma(\overline{m})$ & Std dev of residuals about unweighted \dotfill & L & 8 \\
    & mean magnitude,~$\overline{m}$\\[3pt]
    $\sigma(R_q)$ & Std dev of residuals from piece- 
    \dotfill & M & 10 \\
    & wise quadratic fits$^{*}$\\
    \bottomrule
\end{tabular}
\caption{Classifier features used to identify new Mira candidates, based on \citet{yuan2017a} and listed according to their rank in that publication.\\ L: light curve; M: model. *: $\sigma(R_q)$ was not used as a stand-alone parameter; it is only described to define $\sigma(R_q) / \sigma(\overline{m})$.}
\label{tab:classifier_params}
\end{table}

\begin{figure}
 \includegraphics[width=0.48\textwidth]{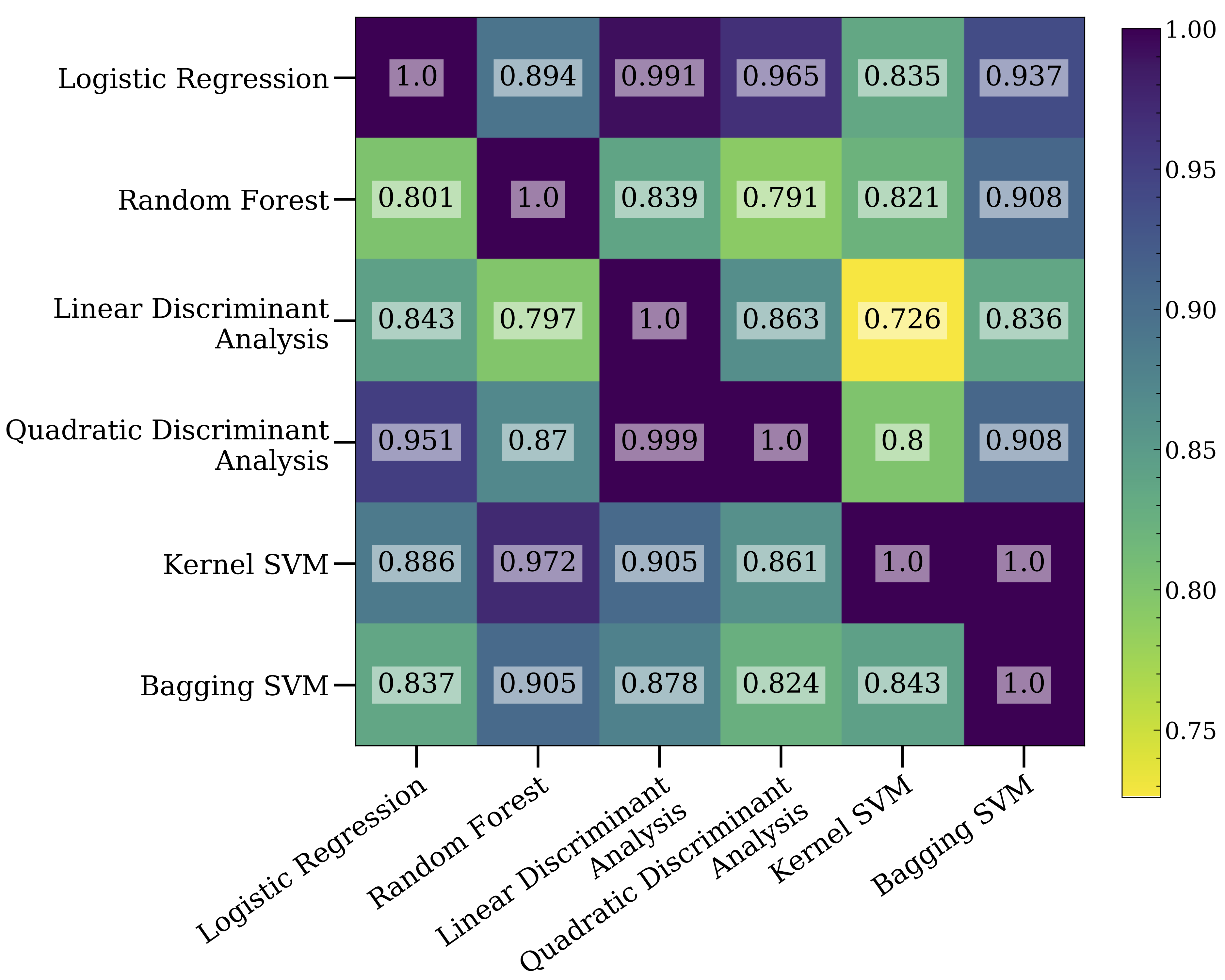}
 \centering
 \caption{Fraction of common candidates across the samples returned by the six machine learning classifiers.}
 \label{fig:classifier_sample_overlap}
\end{figure}

The classifiers were provided with the first five features described in Table~\ref{tab:classifier_params}, which are associated with the $i$-band light curves and best-fit models. These features were amongst those used to identify Mira candidates in \citet{yuan2017a}. We could not include other features from that work because they require periodograms, which our simple sinusoidal fit could not provide. 

We trained and validated the classifiers using the Mira candidates from \citet{yuan2018} that we recovered in our data, as well as the objects that did not pass sample cuts \#1-5 described in \S\ref{sec:mira_id} and Table~\ref{tab:cuts}. The former were considered as known Miras while the latter were considered as known non-Miras.

After training and validating the classifiers, we used the scores assigned to the validation Miras and non-Miras to create Receiver Operating Characteristic (ROC) curves and determine a threshold for each classifier that would separate Miras from non-Miras. The threshold for each method was determined by maximizing the geometric mean, which is defined as $\sqrt{\text{sensitivity} \times \text{specificity}}$. Using the geometric mean to determine the Mira/non-Mira threshold allows for a balance between classifier performance on both the majority and minority classes. It also avoids overfitting the negative class (non-Miras) and under-fitting the positive class (Miras). The Mira/non-Mira thresholds and area under the ROC curve (AUC) for each classifier are shown in Table \ref{tab:classifier_thresholds}. 

We selected a sample of Mira candidates for each classifier by retaining the objects with a classifier score greater than or equal to the respective Mira/non-Mira threshold. We visually inspected the light curves of every variable identified as a Mira by at least one classifier ($\sim$53\% of the remaining sample) and labeled each object as high, low, or no confidence. We retained only the high-confidence objects, which consisted of $\sim$94\% of the remaining objects. Table~\ref{tab:mira_cands_per_classifier} gives the initial number of candidates associated with each classifier and the number that remained after visual inspection. The fractions of initial candidates in common across classifiers are shown in Fig.~\ref{fig:classifier_sample_overlap}. We define ``bronze'', ``silver'' and ``gold'' Mira candidate samples as those identified by one, three or all six classifiers, respectively.

\subsection{P-L relations from the Machine-Learning Samples}

We fit Period-Luminosity relations to various subsets of the 3,052 newly-classified and visually-inspected Mira candidates and the 934 recovered Miras that passed the selection criteria described in \S4. We considered both linear and quadratic relations, defined as:
\begin{equation}\label{eq:plr}
 m \ = \ a_0 \ + \ a_1 (\text{log}_{10} P - 2.3)\ [ \ + \ a_2 (\text{log}_{10} P - 2.3)^2\ ]
\end{equation}
and performed error-weighted fits with iterative 2.5$\sigma$ clipping. The photometric uncertainties were rescaled to obtain $\chi^2_\nu$=1 for each linear fit. The same scaling factor was applied to the corresponding sample prior to performing the equivalent quadratic fit. Results for the latter are only reported if $a_2$ was detected at $\geq$4$\sigma$ significance and its inclusion led to a reduction in $\chi^2_\nu$.

We defined subsamples according to type (O-, C-rich, or all), pulsation mode (fundamental or first overtone), and classifier output. The O/C-rich classification has already been described in \S3.3. We adopted the following dividing lines between first-overtone and fundamental mode pulsators:
 \begin{itemize}
 \item $W_{JK_S}=${16.7}$-6.0\ (\log P-2.2)$;
 \item $W_{JH}={16.7}$-${6.0}$$\ (\log P-2.2)$;
 \item $K_S=17.6-5.0\ (\log P-2.2)$;
 \item $H=17.8-4.5\ (\log P-2.2)$;
 \item $J=18.7-4.5\ (\log P-2.2),$
 \end{itemize}
first classifying objects with $W_{JK_S}$ indices, then those with only $W_{JH}$ information, and so on. In this way, 3,986 variables were classified as either first overtone or fundamental mode, while 2,709 were classified as either O- or C-rich. Figs.~\ref{fig:lpv_updated_pls_o}-\ref{fig:lpv_updated_pls_u} show the resulting P-L relations in the NIR bands and the Wesenheit indices.

We first focused on the $W_{JK_S}$ relation for the fundamental-mode O-rich Miras with $P\!<$400~d, which have been widely used for distance determination and exhibit the tightest scatter \citep{yuan2018,chuang2018,chuang2020}. We performed a series of linear fits using increasingly larger values for the minimum semi-amplitude of the light curves in the $i$-band. The results, shown in Fig.~\ref{fig:pl_ai}, indicate a clear trend in zeropoint and slope until $A_i\sim0.5$~mag. A similar trend was seen for fundamental-mode C-rich Miras, and to some extent in the first-overtone samples, although the smaller size of these samples result in noisier results and the latter do not typically reach such large amplitudes. Thus, for the final fits we required $A_i > 0.5$ and 0.3~mag for fundamental-mode and first-overtone pulsators, respectively.

\begin{figure}
\centering
\includegraphics[width=0.48\textwidth]{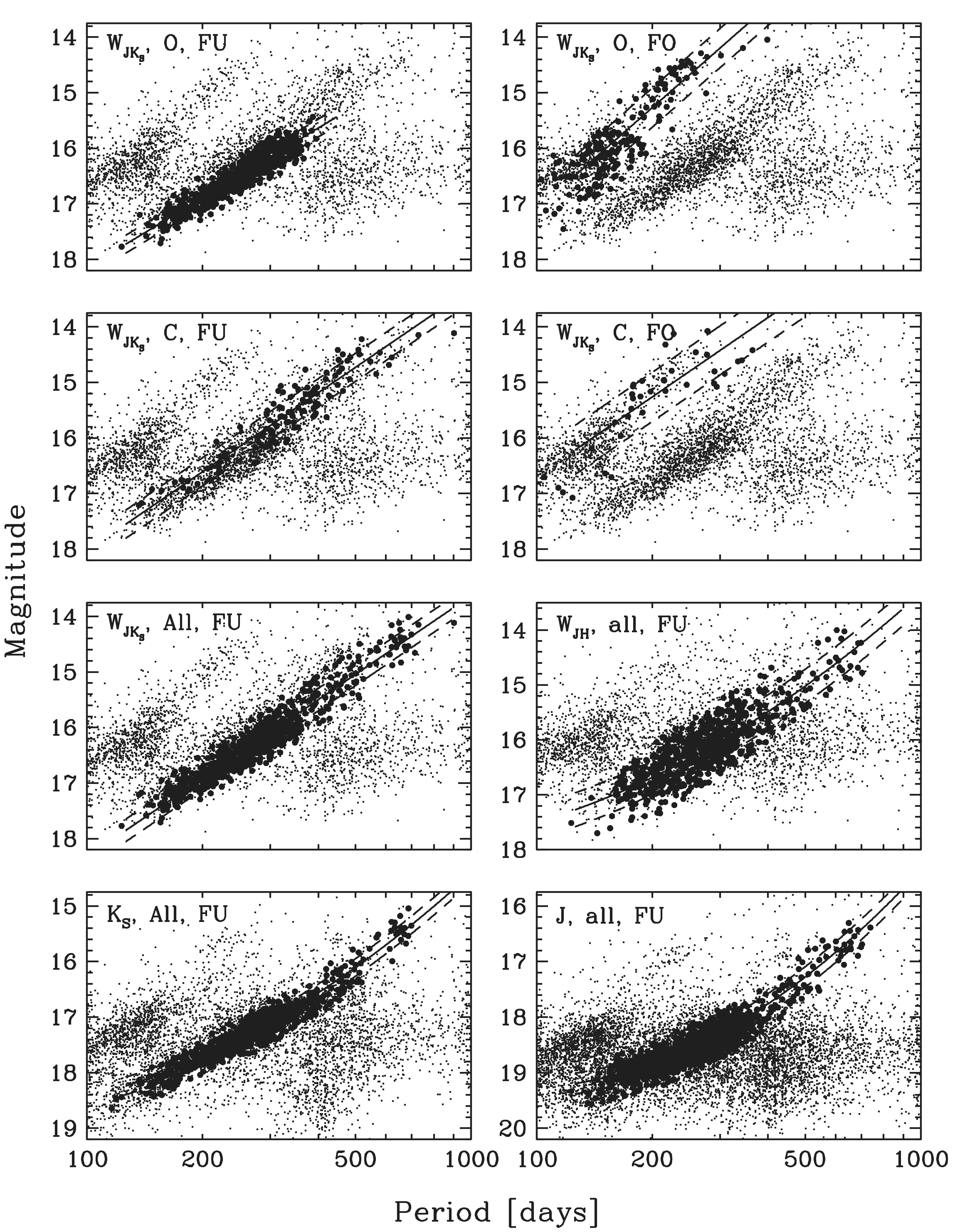}
\caption{Selected P-L relations (solid lines) and $\pm$1$\sigma$ dispersions for various baseline samples.}
\label{fig:classifier_pl_fits_baseline}
\end{figure}

Table~\ref{tab:plr_coefficients} presents the results of all P-L fits, while Fig.~\ref{fig:classifier_pl_fits_baseline} shows selected results for our ``baseline'' samples. In agreement with previous studies, we find the O-rich, fundamental-mode relations exhibit the lowest scatter for a given band or Wesenheit index. In the absence of O/C classification, the resulting fundamental-mode relation usually exhibits a comparable scatter to its O-rich counterpart but it requires a quadratic term. The $K_S$ relations exhibit lower or similar scatter to their $W_{JK_S}$ counterparts, with the $J$ ones being nearly as good. The $H$ relations exhibit slightly worse scatter than the other two bands, while $W_{JH}$ appears markedly worse than the rest.

We do not find a significant improvement when moving from the baseline ``bronze'' sample to the ``silver'' and ``gold'' ones, which is not surprising given the high degree of correlation among the samples identified by the different classifiers.

\subsection{P-L Relations based on OGLE LMC Miras}\label{sec:ogle_m33_plr}

\begin{table*}
 \begin{tabular}{llllrrrrrrrrl}
\toprule
\multicolumn{1}{c}{Sample} & \multicolumn{1}{c}{Mag} & \multicolumn{1}{c}{Type} & \multicolumn{1}{c}{Mode} &
\multicolumn{1}{c}{$a_0$} & \multicolumn{1}{c}{$\sigma(a_0)$} & \multicolumn{1}{c}{$a_1$} & \multicolumn{1}{c}{$\sigma(a_1)$} &
\multicolumn{1}{c}{$a_2$} & \multicolumn{1}{c}{$\sigma(a_2)$} & \multicolumn{1}{c}{$N$}   & \multicolumn{1}{c}{$\sigma$}      & \multicolumn{1}{c}{Notes}\\
 & & & & \multicolumn{2}{c}{[mag]} & \multicolumn{2}{c}{[mag/dex]} & \multicolumn{2}{c}{[mag/dex$^2$]} & & \multicolumn{1}{c}{[mag]} & \\
\midrule
Bronze & $W_{JK_S}$ & O   & FU & 16.895 & 0.012 & -4.173 & 0.079 &   \nd  &  \nd  &  580 & 0.161 & a \\
Bronze & $W_{JK_S}$ & O   & FO & 15.274 & 0.026 & -6.076 & 0.225 &   \nd  &  \nd  &  174 & 0.353 &   \\
Bronze & $W_{JK_S}$ & C   & FU & 16.613 & 0.049 & -4.722 & 0.181 &   \nd  &  \nd  &  119 & 0.253 &   \\
Bronze & $W_{JK_S}$ & C   & FO & 15.267 & 0.071 & -4.743 & 0.578 &   \nd  &  \nd  &   36 & 0.442 &   \\
Bronze & $W_{JK_S}$ & All & FU & 16.916 & 0.013 & -4.697 & 0.052 &   \nd  &  \nd  &  742 & 0.202 &   \\
Bronze & $W_{JK_S}$ & All & FO & 15.284 & 0.024 & -5.897 & 0.206 &   \nd  &  \nd  &  210 & 0.372 &   \\
\midrule
Bronze & $W_{JH}$   & O   & FU & 16.723 & 0.023 & -3.802 & 0.143 &   \nd  &  \nd  &  538 & 0.287 & a \\
Bronze & $W_{JH}$   & O   & FO & 15.176 & 0.031 & -4.957 & 0.266 &   \nd  &  \nd  &  172 & 0.430 &   \\
Bronze & $W_{JH}$   & C   & FU & 16.264 & 0.058 & -3.158 & 0.249 &   \nd  &  \nd  &  113 & 0.313 &   \\
Bronze & $W_{JH}$   & C   & FO & 15.215 & 0.069 & -3.433 & 0.488 &   \nd  &  \nd  &   51 & 0.506 &   \\
Bronze & $W_{JH}$   & All & FU & 16.688 & 0.025 & -3.339 & 0.222 & -2.093 & 0.440 &  688 & 0.308 &   \\
Bronze & $W_{JH}$   & All & FO & 15.190 & 0.028 & -4.687 & 0.230 &   \nd  &  \nd  &  223 & 0.448 &   \\
\midrule
Bronze & $K_S$      & O   & FU & 17.711 & 0.010 & -3.670 & 0.066 &   \nd  &  \nd  &  592 & 0.137 & a \\
Bronze & $K_S$      & O   & FO & 16.229 & 0.026 & -5.964 & 0.225 &   \nd  &  \nd  &  174 & 0.324 &   \\
Bronze & $K_S$      & C   & FU & 17.559 & 0.042 & -2.579 & 0.150 &   \nd  &  \nd  &  122 & 0.209 &   \\
Bronze & $K_S$      & C   & FO & 16.812 & 0.062 & -3.305 & 0.467 &   \nd  &  \nd  &   43 & 0.386 &   \\
Bronze & $K_S$      & All & FU & 17.709 & 0.012 & -3.112 & 0.107 & -2.279 & 0.227 &  789 & 0.158 &   \\
Bronze & $K_S$      & All & FO & 16.323 & 0.026 & -5.287 & 0.214 &   \nd  &  \nd  &  211 & 0.354 &   \\
\midrule
Bronze & $H$        & O   & FU & 18.053 & 0.011 & -3.360 & 0.071 &   \nd  &  \nd  &  538 & 0.152 & a \\
Bronze & $H$        & O   & FO & 16.638 & 0.031 & -5.638 & 0.276 &   \nd  &  \nd  &  157 & 0.345 &   \\
Bronze & $H$        & C   & FU & 17.973 & 0.038 & -1.410 & 0.164 &   \nd  &  \nd  &  123 & 0.251 &   \\
Bronze & $H$        & C   & FO & 17.395 & 0.058 & -0.895 & 0.486 &   \nd  &  \nd  &   59 & 0.476 &   \\
Bronze & $H$        & All & FU & 18.070 & 0.010 & -3.355 & 0.052 &   \nd  &  \nd  &  871 & 0.183 &   \\
Bronze & $H$        & All & FO & 16.779 & 0.028 & -4.671 & 0.256 &   \nd  &  \nd  &  237 & 0.369 &   \\
\midrule
Bronze & $J$        & O   & FU & 18.821 & 0.011 & -3.094 & 0.070 &   \nd  &  \nd  &  783 & 0.173 & a \\
Bronze & $J$        & O   & FO & 17.455 & 0.023 & -6.084 & 0.214 &   \nd  &  \nd  &  234 & 0.336 &   \\
Bronze & $J$        & C   & FU & 18.974 & 0.047 & -0.842 & 0.197 &   \nd  &  \nd  &  164 & 0.289 &   \\
Bronze & $J$        & C   & FO & 18.531 & 0.091 & -4.505 & 0.519 &   \nd  &  \nd  &   68 & 0.621 &   \\
Bronze & $J$        & All & FU & 18.839 & 0.013 & -2.493 & 0.120 & -3.565 & 0.241 & 1103 & 0.185 &   \\
Bronze & $J$        & All & FO & 17.598 & 0.022 & -5.098 & 0.169 &   \nd  &  \nd  &  323 & 0.366 &   \\
\midrule                                                                                             
\bottomrule
\end{tabular}

 \caption{P-L relations for various subsamples. Only baseline fits are shown; see machine-readable version for the full list. a: P$<$400~d.}
 \label{tab:plr_coefficients}
\end{table*}

As a cross-check on the results from \S4.2, we carried out a more traditional selection of fundamental-mode Mira candidates by using the LMC dataset previously described in \S3.3. We calculated the Wesenheit indices using Eqn.~\ref{eq:wes} and required a minimum semi-amplitude of $A_i\ge 0.5$~mag to match our M33 threshold (equivalent to OGLE's $\Delta I=1.0$~mag). We note \citet{ogle09} adopted $R^{K_S}_{JK_S} = 0.686$, though one obtains consistent results for relative distance moduli as long as the same $R$ values are adopted for both LMC and M33 samples. We also applied the color-color relations derived in \S\ref{sec:colcol} to the LMC sample for consistency.

We fit linear P-L relations following Eq.~\ref{eq:plr} and applying iterative $2.5\sigma$ clipping to the selected LMC Miras with $P<400$~d. We then solved for the intercept of each corresponding relation for the M33 variables while keeping the slope fixed to the LMC-derived value. The resulting PLRs are shown in Fig.~\ref{fig:m33-lmc_FU_plr}, while the PLR coefficients are presented in Table~\ref{tab:ogle_linear_plr_coefficients}. We find good agreement with the PLR zeropoints from Table~\ref{tab:plr_coefficients}, with differences ranging from $<0.01$~mag to $0.07$~mag.

\subsection{Comparison with Yuan et al.~(2018)}

\citet{yuan2018} derived PLRs for fundamental-mode O-rich Miras in M33 using the same bands and indices as our study. They also derived periods based on multi-band sinusoidal light curve fits ($IJHK_S$, in their case) and separated O- and C-rich Miras using $J-H$ and $H-K_S$ color-color relations instead of the procedure we described in \S\ref{sec:colcol}. Their LMC PLRs were based on the same OGLE-III catalog, but they used NIR magnitudes from \citet{yuan2017b}.

We compared the mean magnitudes for Miras in common between our sample and theirs and found good agreement, with mean error-weighted differences (this work $-$ theirs) of $\Delta J=-0.051$, $\Delta H=-0.044$, $\Delta K_S=-0.021$, $\Delta W_{JK_S}=0.009$~mag and scatter of $\sigma\!\sim$0.2~mag, and $\Delta W_{JH}=-0.044$~mag and scatter of $\sigma\!\sim$0.4~mag, which may be due in part to the different pulsation cycles over which the light curves were sampled by the respective studies.

We applied our selection and fitting procedures to the Mira sample from Table 3 of \citet{yuan2018}, adding our $i$ measurements to carry out the color-color selection described in \S\ref{sec:colcol} and obtain the most similar comparison possible. We find reasonable agreement with our results, with zeropoints typically within $0.1$~mag of ours and slopes mutually consistent at $<2\sigma$. We note that the PLR based on the $W_{JH}$ index and their magnitudes also exhibits increased scatter relative to the other ones. The fits are provided in Appendix Table~\ref{tab:plry18}.

\subsection{P-L Relations in \textit{gri}} \label{sec:gri_pl_diagrams}

\citet{iwanek2021} analyzed the light curves of LMC Miras at optical and IR wavelengths and derived variability amplitude ratios and phase lags for different bands. They also generated spectral energy distributions (SEDs) based on a high-quality sample of O- and C-rich Miras. These SEDs were used to create synthetic linear PLRs for O- and C-rich Miras in 42 optical and infrared bands, including $gri$.

We fit $gri$ linear PLRs with $2.5\sigma$ clipping in the form of Eq.~\ref{eq:plr} to the unique O-rich fundamental-mode Mira candidates identified in \S\ref{sec:ml_classifiers} and \ref{sec:ogle_m33_plr} (hereafter referred to as the M33-ML and the M33-LMC samples, respectively). We carried out two sets of fits; one in which we allowed both the intercept and slope to vary and one with slopes fixed to the values from \citet{iwanek2021}. The PLRs are shown in Fig.~\ref{fig:gri_plrs} and the best-fit parameters are listed in Table \ref{tab:gri_params}. The intercepts for candidates from the same sample are consistent within their respective uncertainties while the slopes vary significantly. The $i$-band PLR shows the lowest scatter, though it also was fit using more objects than the $g$- and $r$-band ones.

\subsection{A Mira-based distance to M33}

We estimate a Mira-based distance to M33 using the $W_{JK_S}$ PLRs for fundamental-mode O-rich Miras with $P<400$~d from \S\ref{sec:ml_classifiers} and \ref{sec:ogle_m33_plr}:
\begin{equation}
\mu_{M33} = a_0(M33) - a_0(LMC) + \mu_{LMC}
\end{equation}
where $\mu_{LMC}$ is the LMC distance modulus based on detached eclipsing binaries from \citet{pietrzynski2019}, $18.477\pm0.026$~mag, $a_0(M33)=16.895\pm0.012$~mag comes from the first line of Table~\ref{tab:plr_coefficients}, and $a_0(LMC)=10.743\pm0.020$~mag comes from the first line of Table~\ref{tab:ogle_linear_plr_coefficients}. 

We obtain $\mu_{M33}=24.629\pm0.046$~mag, in good agreement with previous determinations based on a variety of distance indicators \citep[see][and references therein]{breuval23}. The quoted uncertainty is the quadrature sum of the uncertainties listed above, plus an additional 0.03~mag to account for possible systematics in the relative photometric calibration of the LMC and M33 samples.

\section{Conclusions}\label{sec:conclusions}

We used multiband observations to identify over 13,000 new Mira candidates and LPVs in M33. We showed that Mira candidates can be robustly identified by using optical light curves and machine-learning techniques, and our measurements in SDSS bands can be used to guide Mira searches in the Rubin/LSST era.

We use near-infrared measurements to further confirm Mira candidates and classify them into various subsamples, detecting for the first time a clear first-overtone pulsation sequence in this galaxy. We also show that NIR observations are very relevant to creating high-fidelity samples of Miras for distance measurements. We use O-rich fundamental-mode Miras with $P<400$~d to determine a distance modulus for M33 of $\mu=24.629\pm0.046$~mag.

\section*{Acknowledgments}

TK and LMM acknowledge support from the Mitchell Institute for Fundamental Physics \& Astronomy and the Department of Physics \& Astronomy at Texas A\&M University. TK would like to thank Sarah Cantu and Peter Ferguson for many useful discussions, and Taylor Hutchison for input on figures. LM thanks Adam Riess, Wenlong Yuan, Richard Anderson and Martin Groenewegen for helpful suggestions and comments. We thank the anonymous reviewer for their valuable recommendations. JH's contributions were partially done while he was the Arseven/Mitchell Chair in Astronomical Statistics at Texas A\&M University. 

The authors are solely responsible for the content of this paper, which does not necessarily represent the views of the National Academies of Sciences, Engineering, and Medicine.

The authors acknowledge the Texas A\&M University Brazos HPC cluster that contributed to the research reported here. This research was also supported by the Munich Institute for Astro-, Particle and BioPhysics (MIAPbP) which is funded by the Deutsche Forschungsgemeinschaft (DFG, German Research Foundation) under Germany's Excellence Strategy -- EXC-2094 -- 390783311. This research was also supported by the International Space Science Institute (ISSI) in Bern, through ISSI International Team project \#490, ``SHoT: The Stellar Path to the H$_0$ Tension in the Gaia, TESS, LSST and JWST Era.''

This work is based on observations obtained with MegaPrime/ MegaCam, a joint project of CFHT and CEA/DAPNIA, at CFHT which is operated by the National Research Council (NRC) of Canada, the Institut National des Science de l'Univers of the Centre National de la Recherche Scientifique (CNRS) of France, and the University of Hawai\okina i, and the Wide-field InfraRed Camera, a joint project of CFHT, Taiwan, Korea, Canada, France, at CFHT which is operated by the NRC of Canada, the CNRS of France, and the University of Hawai\okina i. This research used the facilities of the Canadian Astronomy Data Centre operated by the National Research Council of Canada with the support of the Canadian Space Agency. 

The Pan-STARRS1 Surveys (PS1) and the PS1 public science archive have been made possible through contributions by the Institute for Astronomy, the University of Hawai\okina i, the Pan-STARRS Project Office, the Max-Planck Society and its participating institutes, the Max Planck Institute for Astronomy, Heidelberg and the Max Planck Institute for Extraterrestrial Physics, Garching, The Johns Hopkins University, Durham University, the University of Edinburgh, the Queen's University Belfast, the Harvard-Smithsonian Center for Astrophysics, the Las Cumbres Observatory Global Telescope Network Incorporated, the National Central University of Taiwan, the Space Telescope Science Institute, the National Aeronautics and Space Administration under Grant No. NNX08AR22G issued through the Planetary Science Division of the NASA Science Mission Directorate, the National Science Foundation Grant No. AST-1238877, the University of Maryland, Eotvos Lorand University (ELTE), the Los Alamos National Laboratory, and the Gordon and Betty Moore Foundation.

\noindent \textit{Facilities}: CFHT

\noindent \textit{Software}:
 Brazos Computational Resource; 
 {\fontfamily{pcr}\selectfont DAOPHOT}, {\fontfamily{pcr}\selectfont ALLSTAR}, {\fontfamily{pcr}\selectfont ALLFRAME}, {\fontfamily{pcr}\selectfont TRIAL} (\citealt{stetson1987}, \citealt{stetson1994}, \citealt{stetson1996});
 Astropy \citep{astropy1,astropy2}; 
 Project Jupyter \citep{jupyter}; 
 Matplotlib \citep{matplotlib}; 
 Numpy \citep{numpy}; 
 Pandas \citep{pandas}; 
 SciPy \citep{scipy};
 IRAF \citep{iraf};
 SAOImage DS9 \citep{saods9}.

\section*{Data Availability}

The data underlying this article are available in the article, in its online supplementary material, and on the web at \href{http://github.com/lmacri/m33sss\_miras}{github.com/lmacri/m33sss\_miras}.

\clearpage

\bibliographystyle{mnras}
\bibliography{m33miras}
\label{lastpage}

\appendix

\setcounter{table}{0}
\renewcommand{\thetable}{A\arabic{table}}
\setcounter{figure}{0}
\renewcommand{\thefigure}{A\arabic{figure}}

\clearpage

\begin{figure*}
\begin{center}
\includegraphics[width=0.85\textwidth]{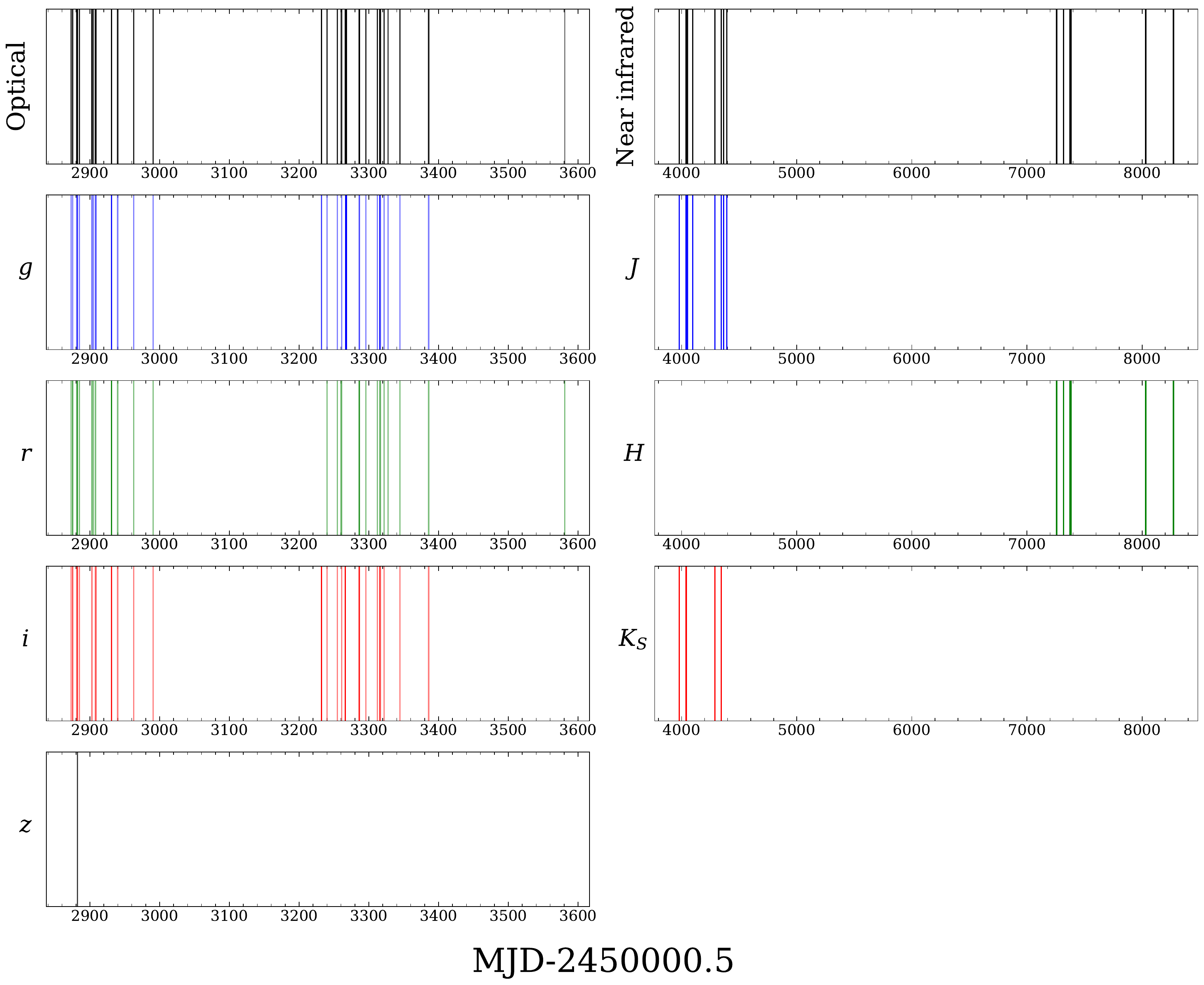}
\end{center}
\caption{Observations across all optical (top left) and NIR bands (top right), \textit{griz} (left, top to bottom), and \textit{JHK$_S$} (right, top to bottom).\label{fig:cadence}}
\end{figure*}

\begin{figure*}
\begin{center}
\includegraphics[width=0.8\textwidth]{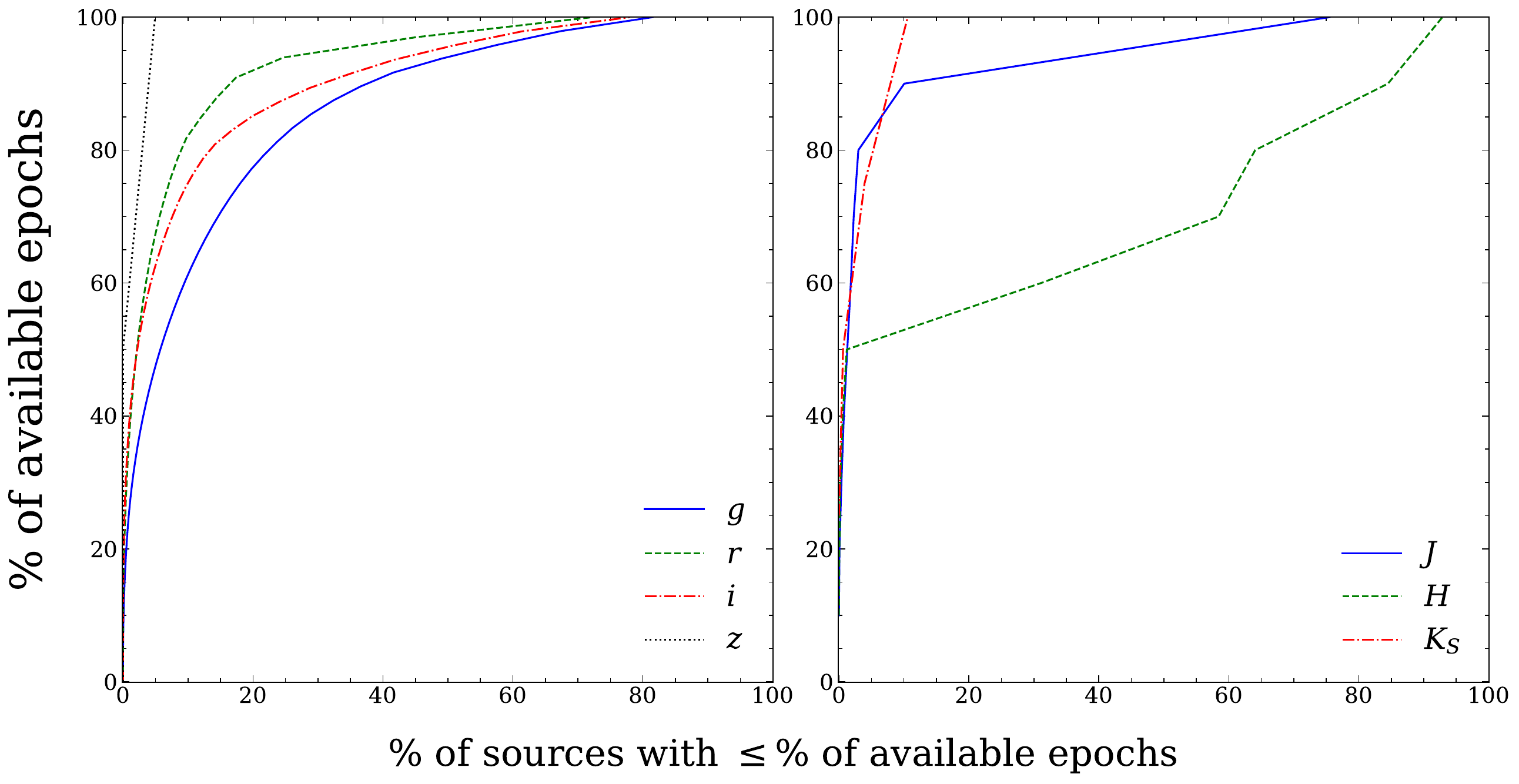}
\end{center}
\caption{Availability of epochs for sources in \textit{griz} (left), and \textit{JHK$_S$} (right).\label{fig:source_avail}}
\end{figure*}

\clearpage

\begin{figure}
\begin{center}
\includegraphics[width=0.49\textwidth]{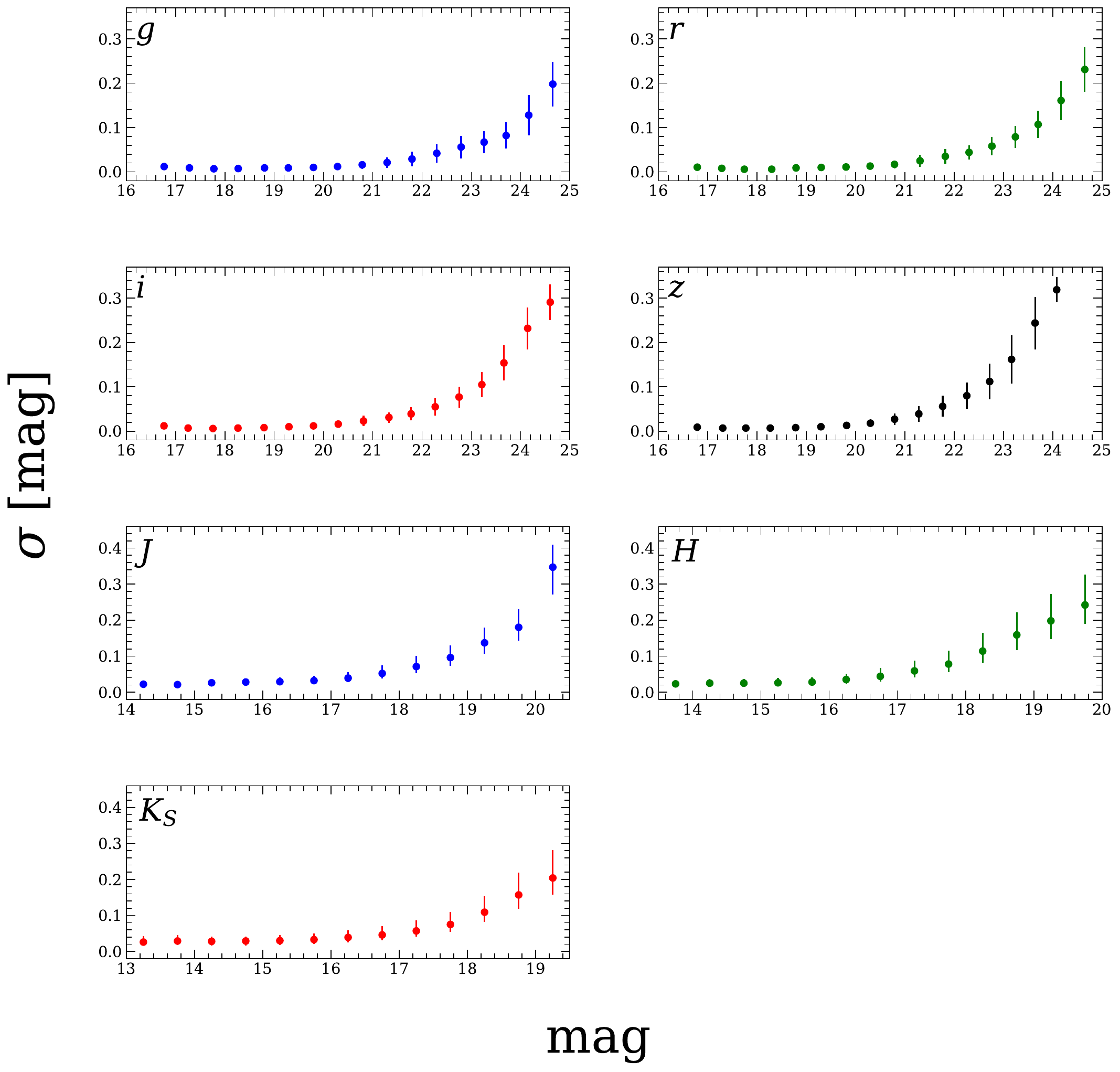}
\end{center}
\caption{Photometric uncertainties as a function of magnitude, binned in 0.5 mag increments.\label{fig:phot_uncer}}
\end{figure}

\begin{table}
\vspace{48pt}
 \centering
 \begin{tabular}{ccrrc}
    \toprule
    Band & Color & \multicolumn{1}{c}{$\chi$} & \multicolumn{1}{c}{$\xi$} & Pivot \\
         &       & \multicolumn{1}{c}{[mag]} & \multicolumn{1}{c}{[mag/mag]} & \multicolumn{1}{c}{[mag]} \\
    \midrule
    $g$ & $g-r$ & $-0.078\pm0.001$ & $0.012\pm0.002$ & 0.8 \\  
        & $g-i$ & $-0.077\pm0.001$ & $0.006\pm0.001$ & 1.5 \\  
        & $g-z$ & $-0.083\pm0.001$ & $0.005\pm0.001$ & 1.5 \\
    \midrule
    $r$ & $g-r$ & $-0.010\pm0.001$ & $0.007\pm0.002$ & 0.8 \\ 
        & $r-i$ & $-0.007\pm0.001$ & $0.026\pm0.002$ & 0.8 \\  
        & $r-z$ & $-0.012\pm0.001$ & $0.017\pm0.002$ & 1.0 \\
    \midrule
    $i$ & $g-i$ & $-0.155\pm0.001$ & $0.023\pm0.001$ & 1.5 \\ 
        & $r-i$ & $-0.155\pm0.001$ & $0.051\pm0.002$ & 0.8 \\  
        & $i-z$ & $-0.151\pm0.001$ & $0.174\pm0.007$ & 0.3 \\
    \midrule
    $z$ & $g-z$ & $0.044\pm0.001$  & $0.011\pm0.001$ & 1.5 \\ 
        & $r-z$ & $0.041\pm0.001$  & $0.013\pm0.002$ & 1.0 \\  
        & $i-z$ & $0.036\pm0.001$  &$-0.012\pm0.006$ & 0.3 \\
    \bottomrule
\end{tabular}
 \caption{$griz$ photometric transformations.}
 \label{tab:pan_trans_coeff}
\end{table}

\begin{table}
\vspace{48pt}
 \centering
 \begin{tabular}{crrr}
\toprule
Band & \multicolumn{1}{c}{$\chi$} & \multicolumn{1}{c}{$\xi$}     & \multicolumn{1}{c}{$\xi^{'}$}\\
     & \multicolumn{1}{c}{[mag]}  & \multicolumn{1}{c}{[mag/mag]} & \multicolumn{1}{c}{[mag/mag$^2$]} \\
\midrule
$J$  & $-0.028\pm0.001$ & $-0.016\pm0.002$ & $-0.035\pm0.004$ \\
$H$  & $-0.043\pm0.001$ & $-0.036\pm0.003$ & $-0.011\pm0.004$ \\
$K_S$& $-0.051\pm0.002$ & $-0.004\pm0.009$ & \multicolumn{1}{c}{\nd} \\
\bottomrule
\end{tabular}

 \caption{$JHK_S$ photometric transformations}
 \label{tab:nir_trans_coeff}
\end{table}

\begin{figure}
\begin{center}
\includegraphics[width=0.49\textwidth]{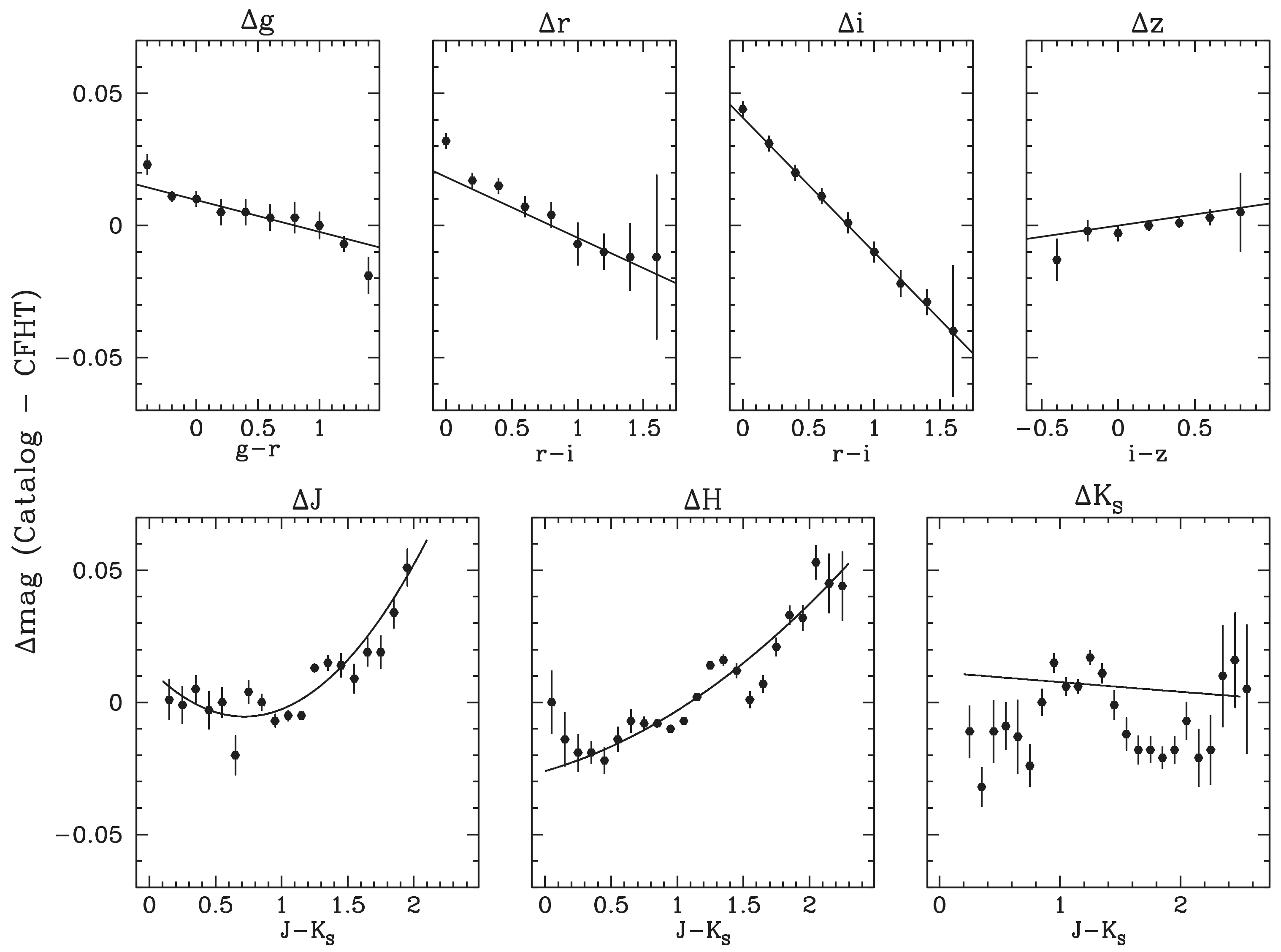}
\end{center}
\caption{Representative photometric transformations for all bands. Symbols are binned residuals while solid lines represent the best-fit color terms.\label{fig:trans_resids}}
\end{figure}

\begin{figure}
\begin{center}
 \includegraphics[width=0.49\textwidth]{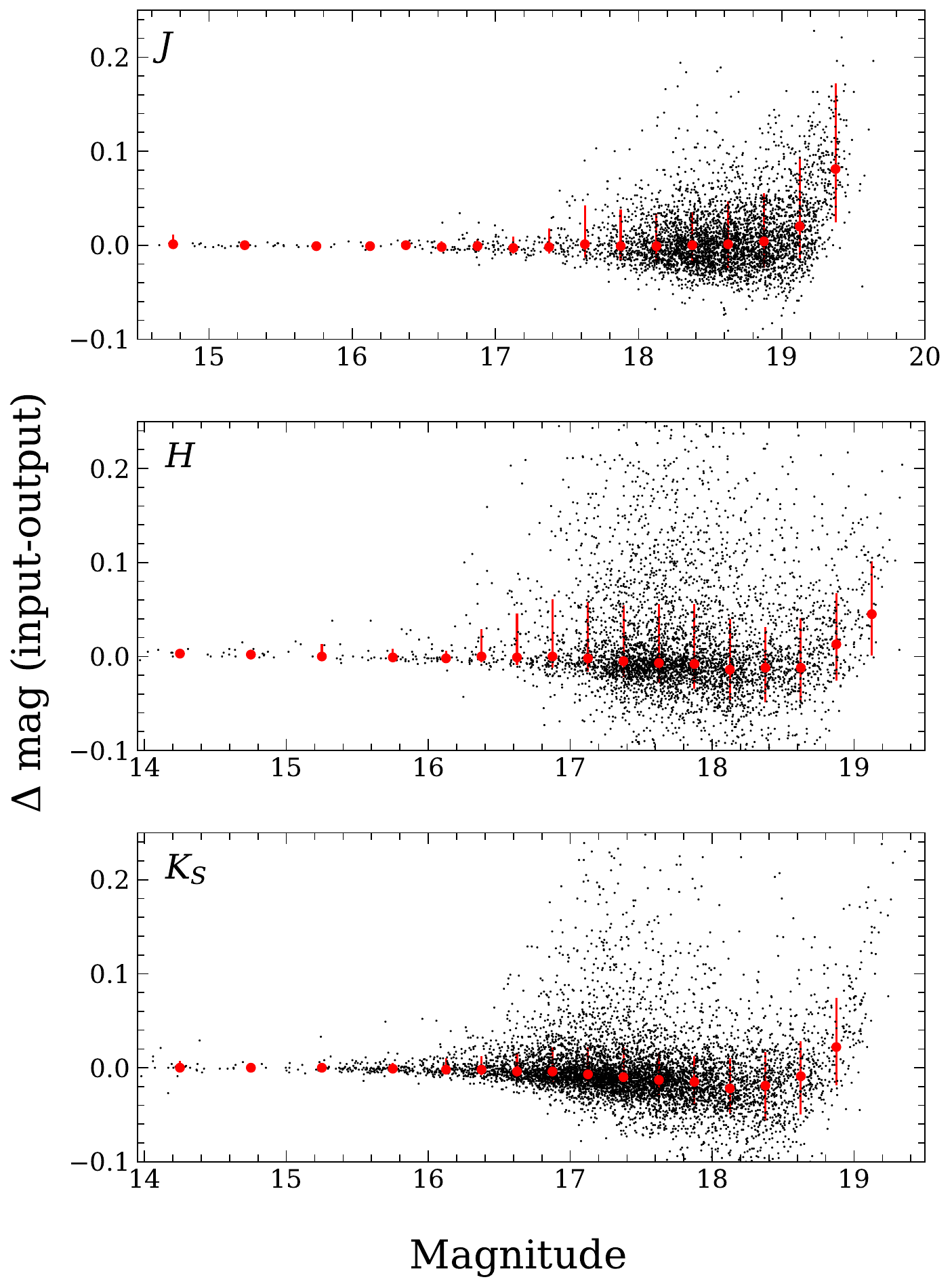}
\end{center}
 \caption{Results of artificial star tests to characterize the crowding bias in the NIR photometry of LPVs and Mira candidates. Small black dots represent the error-weighted mean crowding correction for each object. Larger red filled symbols show median values and red errorbars depict $\pm 1\sigma$ ranges for stars in each bin.\label{fig:crowd}}
\end{figure}

\clearpage

\begin{figure}
\begin{center}
\includegraphics[width=0.48\textwidth]{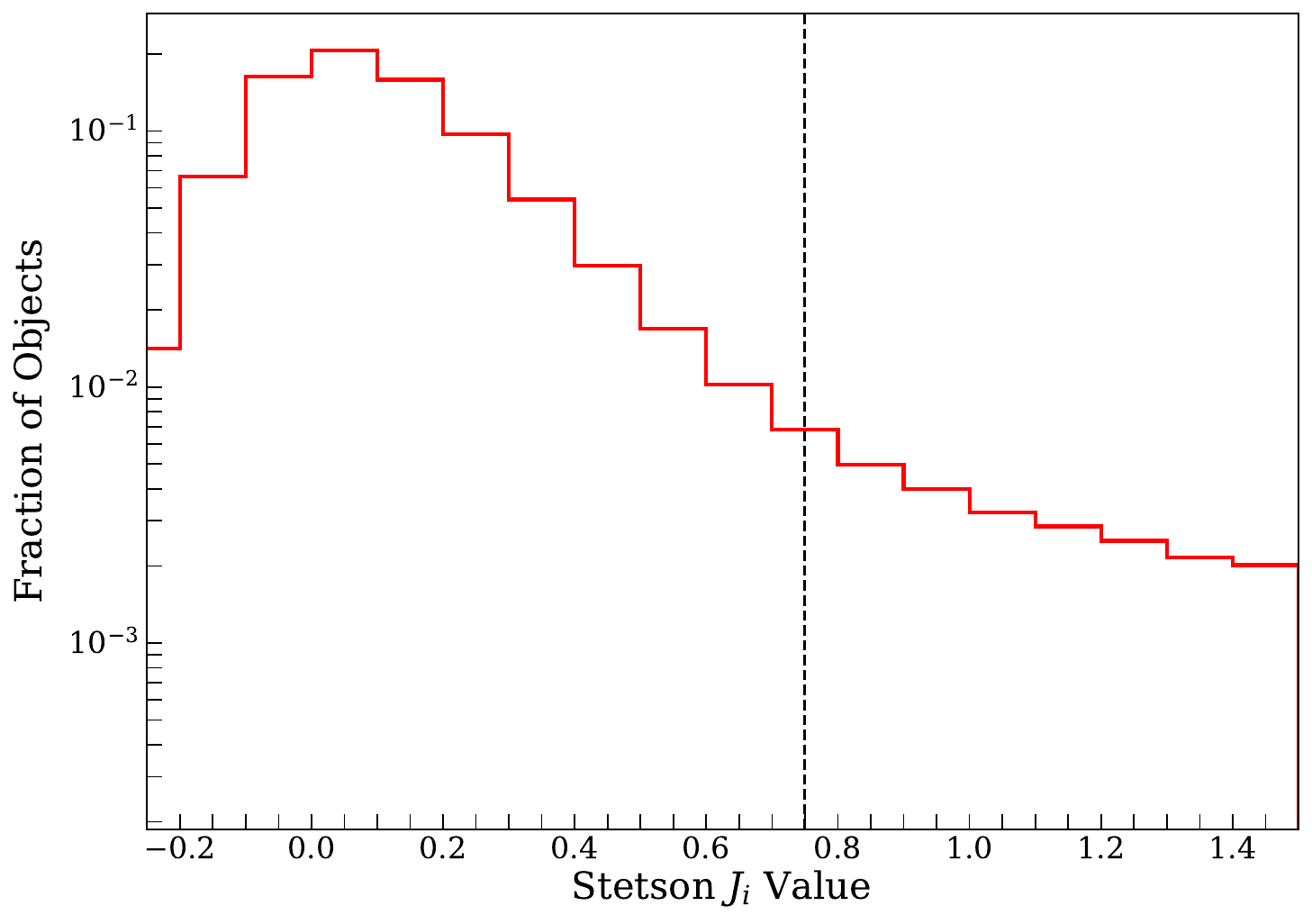}
\end{center}
\caption{Histogram of Stetson $J$ variability index values for all sources detected in $i$ (step 2 of Table~\ref{tab:cuts}). The vertical dashed line at $J_i = 0.75$ is the threshold we adopted.}
\label{fig:ji_hist}
\end{figure}

\begin{figure}
\begin{center}
\includegraphics[width=0.48\textwidth]{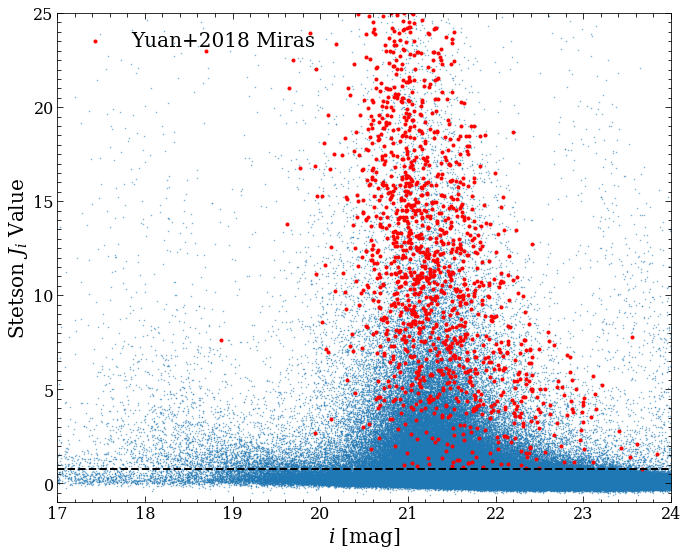}
\end{center}
\caption{$J_i$ versus $i$ for all sources detected in that band (step 2 in Table~\ref{tab:cuts}). The horizontal dashed lined at $J_i = 0.75$ shows the adopted threshold. Recovered Miras from \citet{yuan2018} are overplotted in red.}
\label{fig:ji_vs_i}
\end{figure}

\begin{table}
 \centering
 \begin{tabular}{lr}
\toprule
Criterion & \multicolumn{1}{c}{N}\\
\midrule
1. Detected in $\ge$1 of $gri$ & 1,158,951 \\
2. Detected in $i$ & 1,036,491\\
3. $J_i \geq 0.75$ & 69,798 \\
4. $r\!-\!i\geq 0$ or no $r$ & 65,609 \\
5. $R_i \geq 0.3$~mag & 39,660 \\
6. Detected in $\ge$1 of $JHK_S$ & 14,312 \\
\bottomrule
\end{tabular}

 \caption{Selection of LPVs and Mira candidates.}
 \label{tab:cuts}
\end{table}

\begin{figure}
\begin{center}
\includegraphics[width=0.48\textwidth]{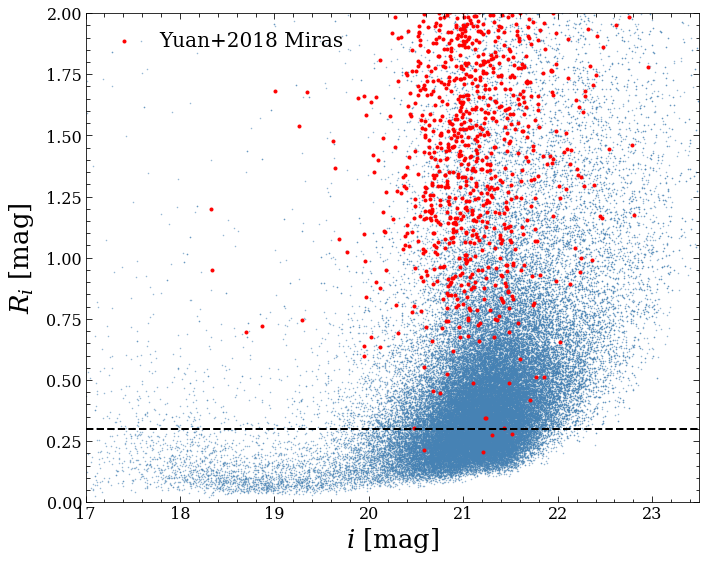}
\end{center}
\caption{Range of magnitudes spanned by $i$ light curves ($R_i$) versus mean $i$ magnitude for objects with: (i) $J_i \geq 0.75$, (ii) $r-i \geq 0$ or a non-detection in $r$. The horizontal dashed line shows our adopted threshold of $R_i \geq 0.3$. The Miras from \citet{yuan2018} are overplotted in red.}
\label{fig:ai_hist}
\end{figure}

\begin{figure}
\begin{center}
\includegraphics[width=0.48\textwidth]{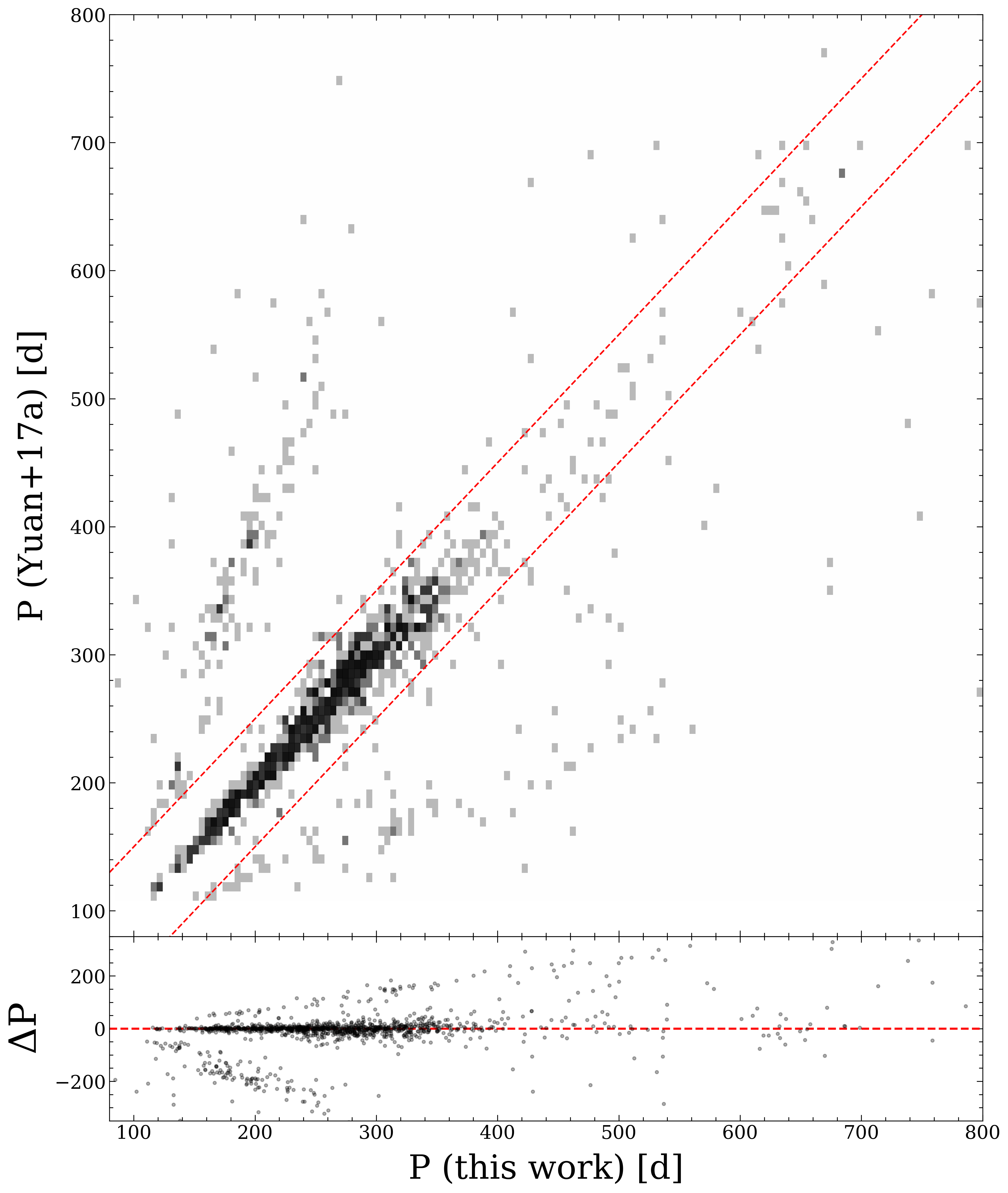}
\end{center}
\caption{Top: Comparison of periods for Mira candidates from \citet{yuan2017a} that were recovered in this work; $\sim$78.1\% of them had $\Delta P/P<50$~d. Bottom: Residuals from the 1:1 relation.}
\label{fig:compare_y17a_periods}
\end{figure}

\clearpage
\begin{table*}
 \centering
 \tiny{
\setlength{\tabcolsep}{4pt}
\begin{tabular}{lrrrrrrrrrrrrrrrrrrr}
\toprule
ID & \multicolumn{1}{c}{RA} & \multicolumn{1}{c}{Dec} & \multicolumn{1}{c}{$i$} & \multicolumn{1}{c}{$\sigma_{i}$} & \multicolumn{1}{c}{$r$} & \multicolumn{1}{c}{$\sigma_{r}$} & \multicolumn{1}{c}{$g$} & \multicolumn{1}{c}{$\sigma_{g}$} & \multicolumn{1}{c}{$z$} & \multicolumn{1}{c}{$\sigma_{z}$} & \multicolumn{1}{c}{$J$} & \multicolumn{1}{c}{$\sigma_{J}$} & \multicolumn{1}{c}{$H$} & \multicolumn{1}{c}{$\sigma_{H}$} & \multicolumn{1}{c}{$K_S$} & \multicolumn{1}{c}{$\sigma_{K_S}$} & \multicolumn{1}{c}{$J_i$} & \multicolumn{1}{c}{$J_r$} & \multicolumn{1}{c}{$J_g$}\\
& \multicolumn{2}{c}{[deg]} & \multicolumn{2}{c}{[mag]} &  \multicolumn{2}{c}{[mag]} &  \multicolumn{2}{c}{[mag]} &  \multicolumn{2}{c}{[mag]} &  \multicolumn{2}{c}{[mag]} &  \multicolumn{2}{c}{[mag]} &  \multicolumn{2}{c}{[mag]} &  \multicolumn{3}{c}{} \\
\midrule
01331383+3036355 & 23.307608 & 30.609863 & 20.530 & 0.013 & 22.692 & 0.024 & 24.635 & 0.046 & 19.773 & 0.006 & 17.788 & 0.007 & 16.969 & 0.010 & 16.531 & 0.008 &  4.245 &  1.342 &  0.408\\
01331387+3035040 & 23.307783 & 30.584454 & 21.093 & 0.004 & 22.068 & 0.006 & 24.002 & 0.025 & 20.682 & 0.011 & 19.179 & 0.020 & 18.375 & 0.016 & 17.981 & 0.011 &  0.627 & -0.014 &  0.290\\
01331390+3035156 & 23.307920 & 30.587658 & 20.781 & 0.003 & 21.644 & 0.007 & 22.744 & 0.009 & 20.419 & 0.018 & 18.976 & 0.015 & 18.133 & 0.013 & 17.822 & 0.010 &  0.394 &  0.436 &  0.463\\
01331393+3035343 & 23.308054 & 30.592873 & 21.054 & 0.003 & 21.587 & 0.005 & 22.587 & 0.005 & 20.832 & 0.014 & 19.359 & 0.056 & 18.631 & 0.019 & 18.521 & 0.018 & -0.112 &  0.230 &  0.026\\
01331398+3036270 & 23.308245 & 30.607506 & 18.802 & 0.007 & 19.745 & 0.015 & 21.123 & 0.007 & 18.524 & 0.007 & 16.819 & 0.006 & 16.050 & 0.004 & 15.699 & 0.006 &  3.553 &  8.391 &  1.476\\
01331400+3036586 & 23.308348 & 30.616270 & 20.356 & 0.003 & 21.617 & 0.006 & 23.127 & 0.010 & 19.904 & 0.010 & 18.249 & 0.011 & 17.408 & 0.009 & 17.070 & 0.008 &  0.144 &  0.501 & -0.000\\
01331401+3036464 & 23.308376 & 30.612894 & 19.221 & 0.002 & 19.925 & 0.004 & 21.189 & 0.005 & 18.915 & 0.004 & 17.391 & 0.006 & 16.636 & 0.006 & 16.329 & 0.006 &  0.637 &  1.796 &  0.998\\
01331403+3033509 & 23.308460 & 30.564144 & 21.011 & 0.009 & 22.482 & 0.023 & 24.311 & 0.036 & 20.357 & 0.007 & 18.723 & 0.014 & 17.800 & 0.011 & 17.383 & 0.010 &  2.138 &  1.983 &  0.336\\
\midrule
\bottomrule
\end{tabular}}
 \caption{Calibrated photometry of all point sources detected in our analysis. No crowding corrections have been applied.\\The full version of this table is available online; only a few representative lines are shown here for guidance.}
 \label{tab:calphot}
\end{table*}

\begin{figure*}
\centering \includegraphics[height=0.75\textheight]{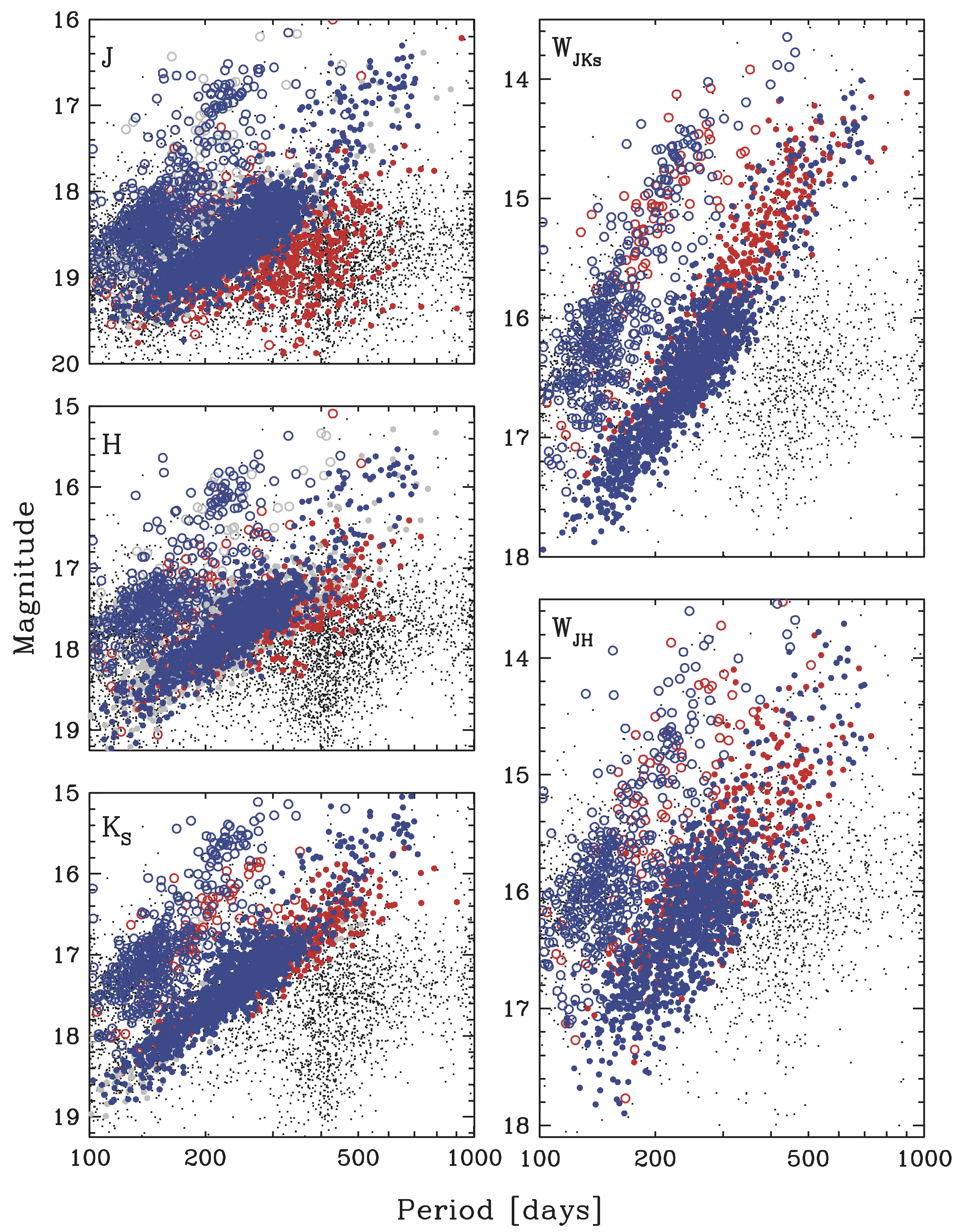}
\caption{P-L relations in NIR bands and Wesenheit indices for the 3,986 Mira candidates described in \S4.2. Variables that did not pass the cuts are plotted using small points. Open and filled circles denote first-overtone and fundamental-mode pulsators. Objects plotted in red and blue were classified as C- or O-rich, respectively, while those in grey only have measurements in one NIR band and thus cannot be classified.}
\label{fig:lpv_updated_pls_o}
\end{figure*}

\clearpage

\begin{figure}
\centering \includegraphics[width=0.49\textwidth]{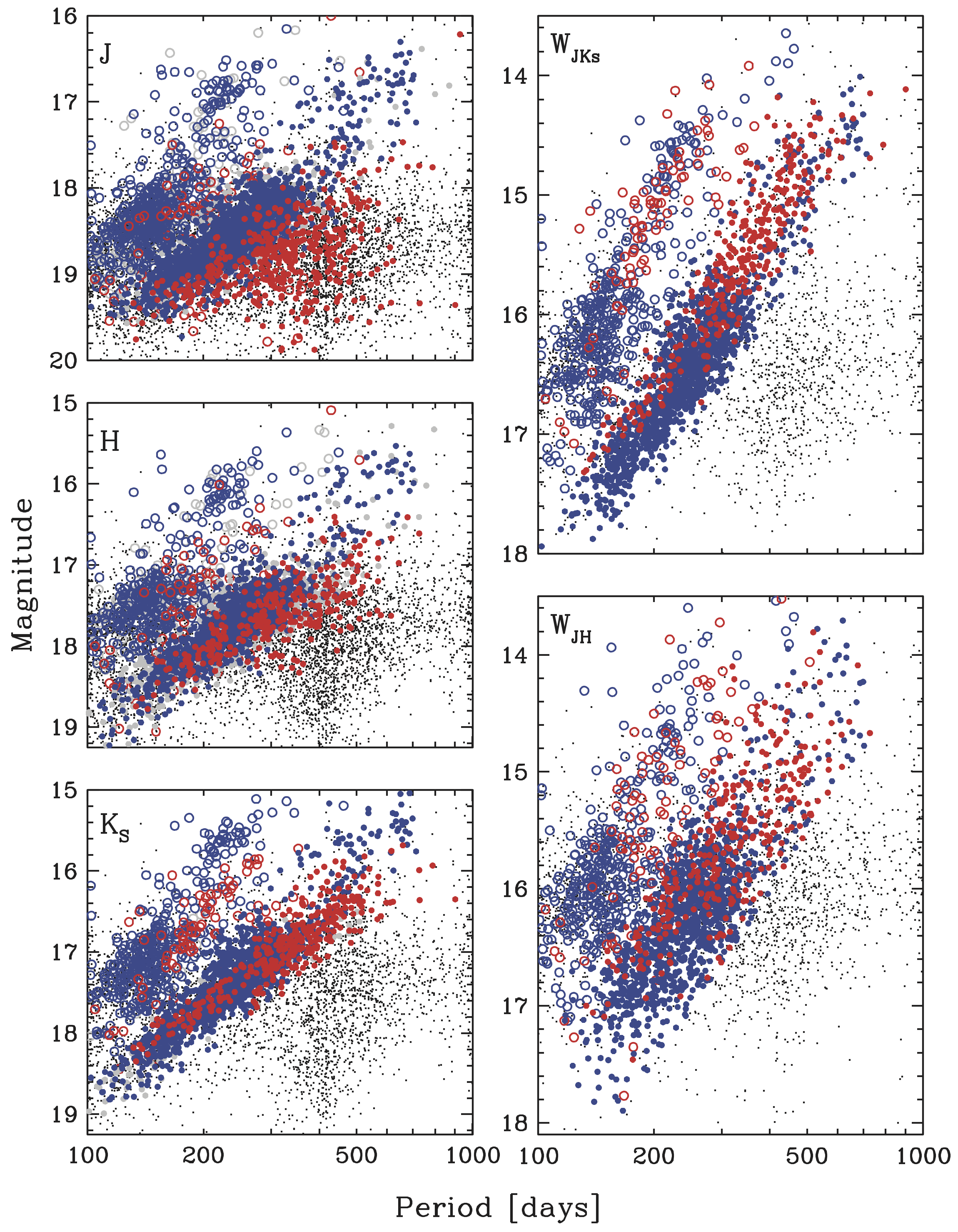}
\caption{Same as \ref{fig:lpv_updated_pls_o}, but focusing on the C-rich LPVs.}
\label{fig:lpv_updated_pls_c}
\end{figure}

\begin{figure}
\centering \includegraphics[width=0.49\textwidth]{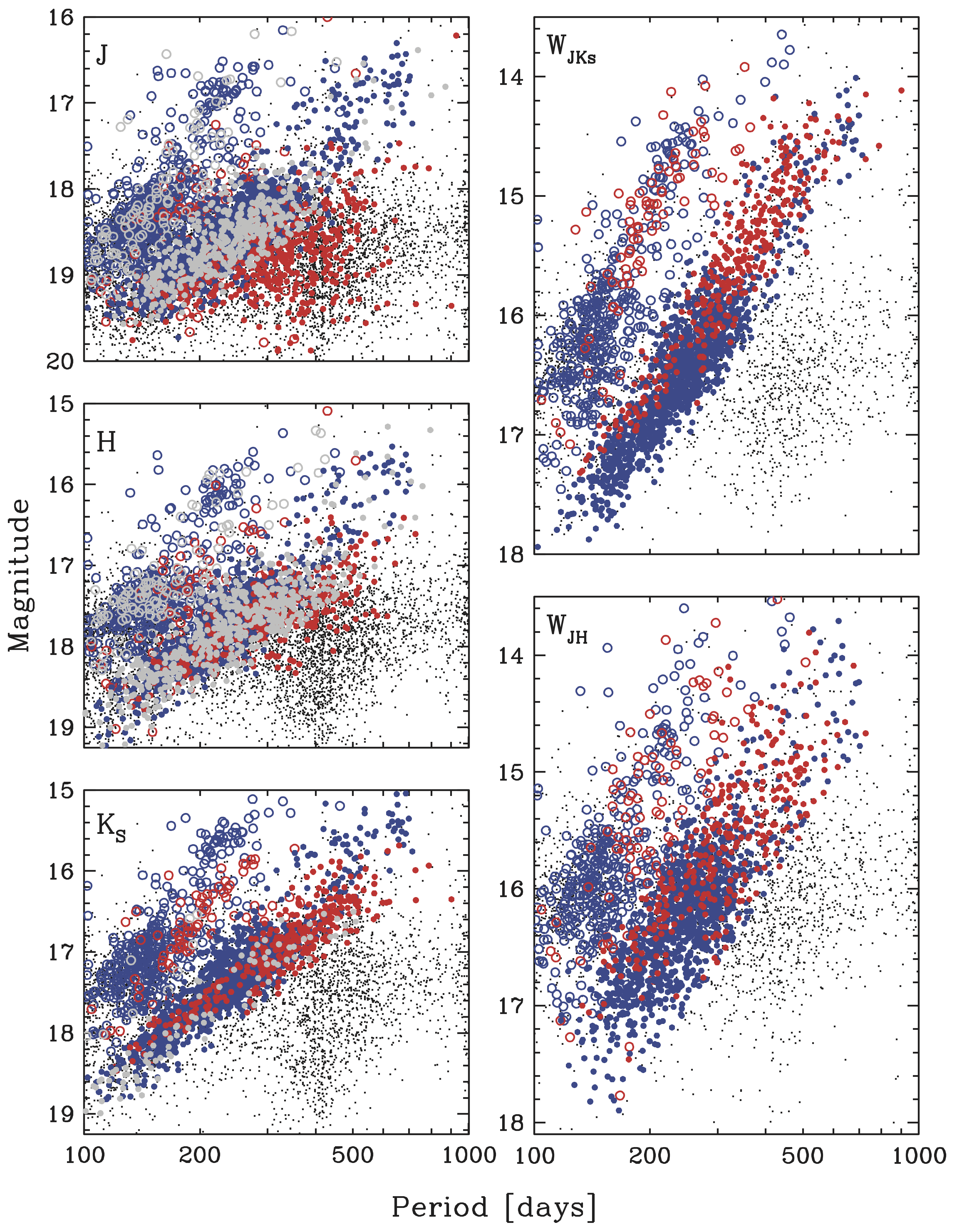}
\caption{Same as \ref{fig:lpv_updated_pls_o}, but focusing on the LPVs without O/C classification.}
\label{fig:lpv_updated_pls_u}
\end{figure}

\begin{figure}
\centering \includegraphics[width=0.48\textwidth]{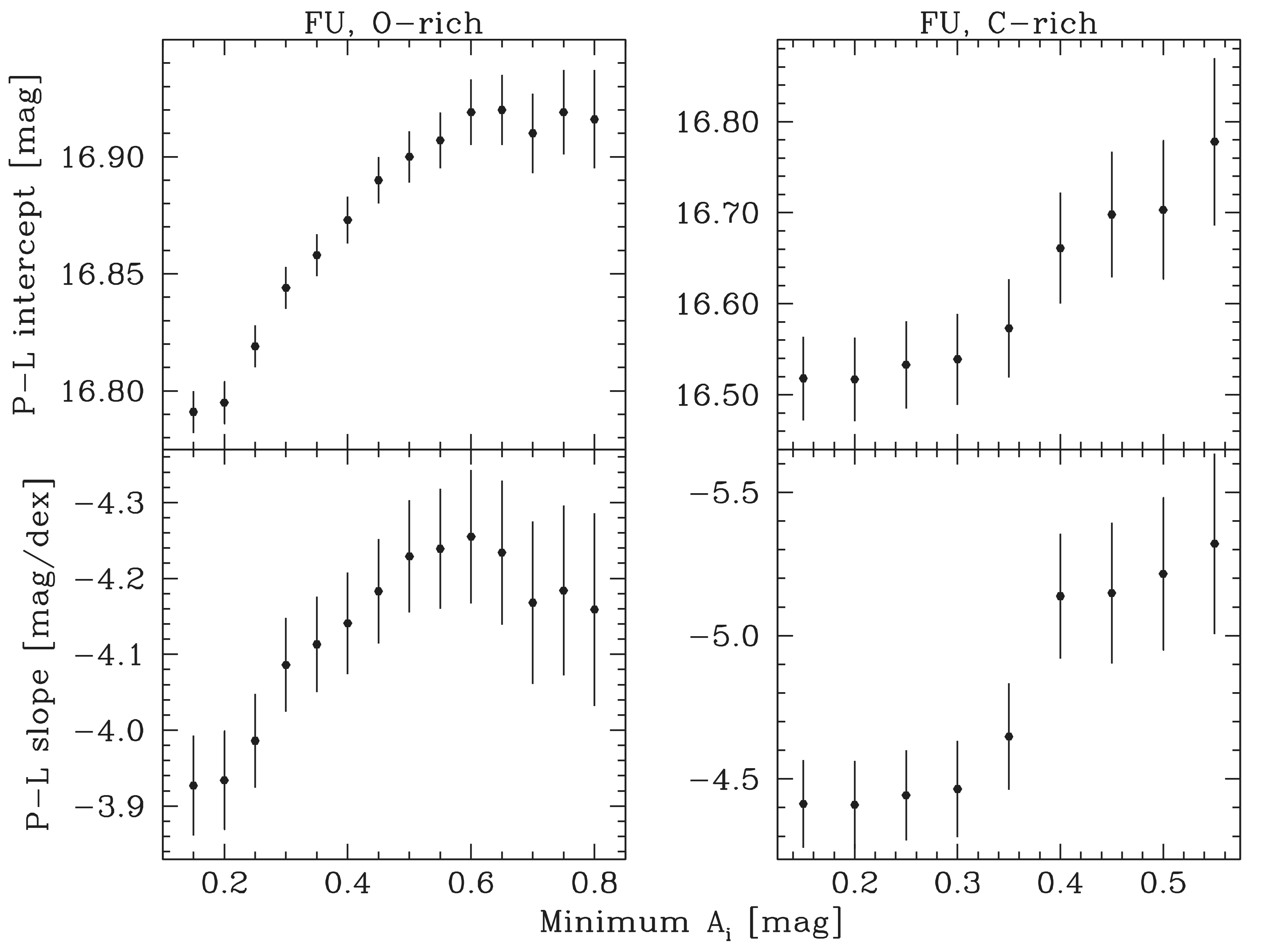}
\caption{Result of linear fits to the $W_{JK_S}$ P-L relations of fundamental-mode pulsators as a function of minimum $i$-band amplitude. Left: O-rich, $P<$400~d. Right: C-rich. Top: P-L intercepts. Bottom: P-L slopes. We adopt a minimum value of $A_i=0.5$~mag for our final fits.}
 \label{fig:pl_ai}
\end{figure}

\begin{table}
 \centering
 \begin{tabular}{lcc}
    \toprule
    Classifier & Threshold & AUC \\
    \midrule
    Logistic Regression & 0.603 & 0.991 \\ 
    Random Forest & 0.195 & 0.993 \\ 
    Linear Discriminant Analysis & 0.758 & 0.990 \\
    Quadratic Discriminant Analysis & 0.833 & 0.991 \\
    Kernel SVM & 0.845 & 0.993 \\
    Bagging SVM & 0.933 & 0.989 \\
    \bottomrule
\end{tabular}
 \caption{Mira/non-Mira thresholds and AUC values for each classifier.}
 \label{tab:classifier_thresholds}
\end{table}

\begin{table}
 \centering
 \begin{tabular}{lcc}
\toprule
Classifier & Passed & Visual \\
\midrule
Logistic Regression             & 2,335 & 2,206 \\ 
Random Forest                   & 2,533 & 2,404 \\ 
Linear Discriminant Analysis    & 2,800 & 2,637 \\
Quadratic Discriminant Analysis & 2,388 & 2,271 \\
Kernel SVM                      & 2,077 & 1,989 \\
Bagging SVM                     & 2,511 & 2,384 \\
\midrule
Any 1 classifier                & 3,251 & 3,052 \\
At least 3 classifiers          & 2,595 & 2,454 \\
All 6 classifiers               & 1,746 & 1,686 \\
\bottomrule
\end{tabular}

 \caption{Variables passing the threshold for a given classifier, and those remaining after visual inspection, out of a starting sample of 6,144 objects.}
 \label{tab:mira_cands_per_classifier}
\end{table}

\clearpage

\begin{figure*}
 \includegraphics[height=0.85\textheight]{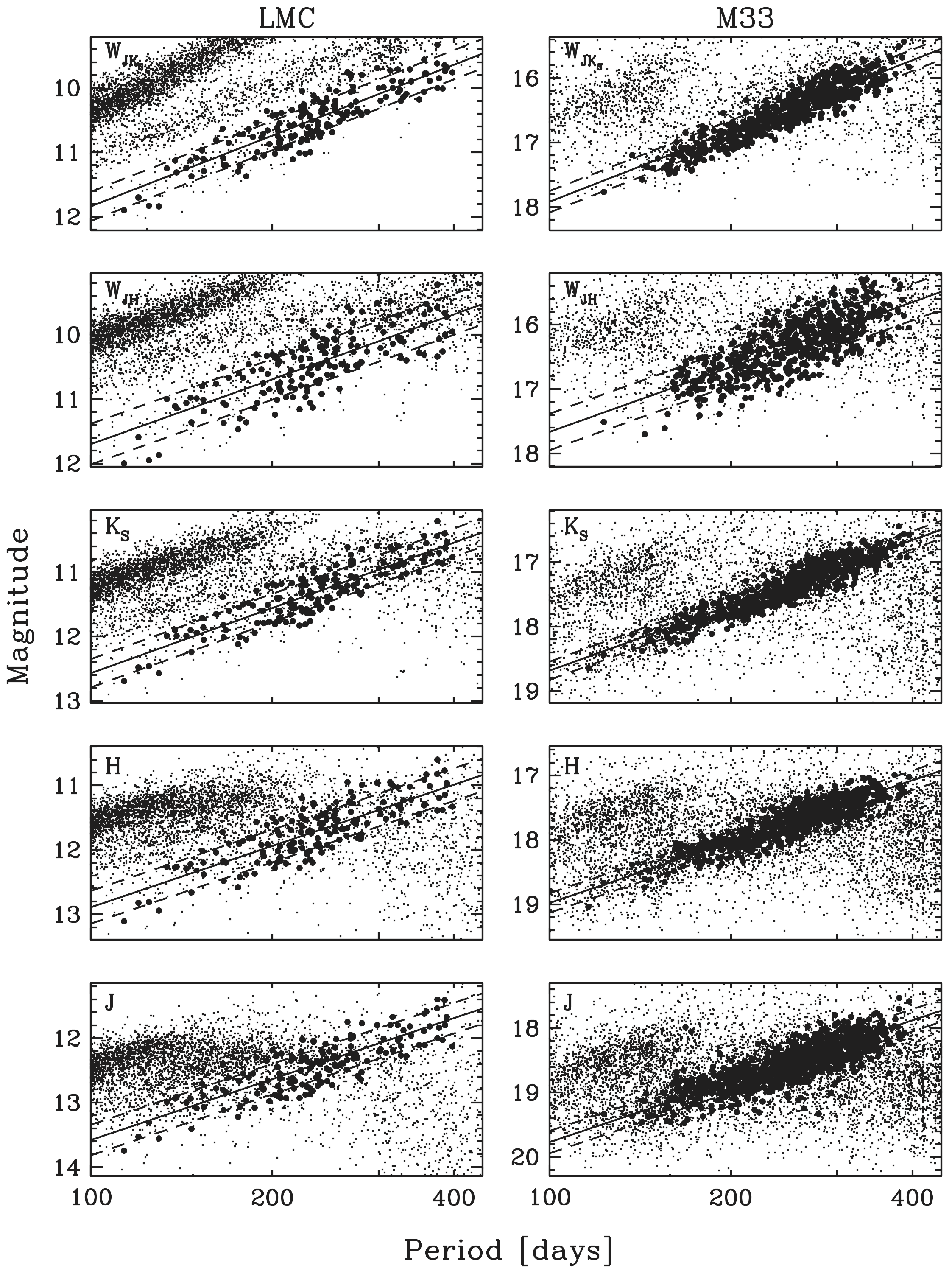}
 \centering
 \caption{P-L relations for fundamental-mode, O-rich Miras (filled symbols) in the LMC (left) and M33 (right) in various Wesenheit indices and bands. The slopes were determined from the LMC samples and fixed for their M33 counterparts. The solid lines indicate the best-fit relations and the dashed lines indicate the $\pm$1$\sigma$ dispersions.}
 \label{fig:m33-lmc_FU_plr}
\end{figure*}

\clearpage

\begin{table*}
 \centering
 \tiny{
\setlength{\tabcolsep}{2pt}
\begin{tabular}{lrrrrrrrrrrrrrrrrrrrrrrrrr}
\toprule
\multicolumn{1}{c}{ID} & \multicolumn{1}{c}{RA} & \multicolumn{1}{c}{Dec} & \multicolumn{1}{c}{$P$} & \multicolumn{1}{c}{$i$} & \multicolumn{1}{c}{$J$} & \multicolumn{1}{c}{$H$} & \multicolumn{1}{c}{$K_S$} & \multicolumn{1}{c}{$r$} & \multicolumn{1}{c}{$g$} & \multicolumn{1}{c}{$A_i$} & \multicolumn{1}{c}{$A_J$} & \multicolumn{1}{c}{$A_r$} & \multicolumn{1}{c}{$A_g$} & \multicolumn{1}{c}{$\phi_i$} & \multicolumn{1}{c}{$\phi_J$} & \multicolumn{1}{c}{$\sigma_{P}$} & \multicolumn{1}{c}{$\sigma_{i}$} & \multicolumn{1}{c}{$\sigma_{J}$} & \multicolumn{1}{c}{$\sigma_{H}$} & \multicolumn{1}{c}{$\sigma_{K_S}$} & \multicolumn{1}{c}{$\sigma_{r}$} & \multicolumn{1}{c}{$\sigma_{g}$} & \multicolumn{1}{c}{$\sigma_{A_i}$} & \multicolumn{1}{c}{$\sigma_{A_J}$} &\multicolumn{1}{c}{$\sigma_{A_r}$}\\
\multicolumn{1}{c}{} & \multicolumn{2}{c}{[deg]} &\multicolumn{1}{c}{[d]} & \multicolumn{6}{c}{[mag]} & \multicolumn{4}{c}{[mag]} & \multicolumn{2}{c}{} & \multicolumn{1}{c}{[d]} & \multicolumn{6}{c}{[mag]} & \multicolumn{3}{c}{[mag]} \\
\midrule
01322479+3047175 & 23.103285 & 30.788191 &  389.05 & 20.348 &    \nd & 18.074 &    \nd & 21.936 & 23.175 & 0.766 & 0.105 & 0.316 & 0.326 &  0.157 &  0.165 &    \nd & 0.004 &   \nd & 0.038 &   \nd & 0.008 & 0.005 &   \nd &   \nd & 0.016\\
01322533+3036126 & 23.105537 & 30.603506 & 1071.52 & 21.644 &    \nd & 17.339 &    \nd & 22.958 &    \nd & 0.194 & 0.204 & 0.266 &   \nd &  0.154 &  0.341 &    \nd & 0.008 &   \nd & 0.059 &   \nd & 0.011 &   \nd & 0.012 & 0.073 &   \nd\\
01322583+3047503 & 23.107637 & 30.797293 & 1071.52 & 20.501 &    \nd & 18.077 &    \nd & 22.071 & 23.478 & 0.935 & 0.142 & 0.452 & 0.508 &  0.023 & -0.208 &    \nd & 0.004 &   \nd & 0.183 &   \nd & 0.005 & 0.007 &   \nd &   \nd &   \nd\\
01322793+3039553 & 23.116392 & 30.665359 &  117.82 & 21.206 &    \nd & 17.756 &    \nd & 23.215 &    \nd & 0.155 & 0.092 & 0.257 &   \nd &  0.046 & -0.195 &    \nd & 0.005 &   \nd & 0.023 &   \nd & 0.012 &   \nd &   \nd &   \nd & 0.026\\
01322879+3049434 & 23.119947 & 30.828728 &  365.78 & 21.066 &    \nd & 17.036 &    \nd & 22.939 &    \nd & 0.813 & 0.125 & 0.453 &   \nd &  0.068 &  0.156 &    \nd & 0.004 &   \nd & 0.016 &   \nd & 0.008 &   \nd &   \nd &   \nd &   \nd\\
01323222+3041360 & 23.134270 & 30.693346 &  361.46 & 21.477 &    \nd & 17.590 &    \nd & 22.586 &    \nd & 0.568 & 0.215 & 0.777 &   \nd & -0.092 &  0.167 &    \nd & 0.004 &   \nd & 0.026 &   \nd & 0.013 &   \nd &   \nd &   \nd & 0.029\\
01323651+3037358 & 23.152111 & 30.626614 &  161.01 & 21.957 &    \nd & 17.388 &    \nd &    \nd &    \nd & 0.305 & 0.027 &   \nd &   \nd &  0.186 & -0.181 &    \nd & 0.007 &   \nd & 0.022 &   \nd &   \nd &   \nd &   \nd &   \nd &   \nd\\
01330333+3038314 & 23.263865 & 30.642057 &  244.74 & 21.063 & 20.012 & 18.492 &    \nd & 21.836 &    \nd & 0.144 & 0.082 & 0.152 &   \nd &  0.478 &  0.500 &    \nd & 0.005 & 0.115 & 0.074 &   \nd & 0.007 &   \nd &   \nd &   \nd & 0.013\\
01330346+3041361 & 23.264431 & 30.693350 &  408.17 & 20.696 & 17.825 & 16.674 &    \nd & 21.723 &    \nd & 0.752 & 0.498 & 0.904 &   \nd &  0.500 &  0.184 &   5.53 & 0.006 & 0.037 & 0.034 &   \nd & 0.007 &   \nd &   \nd & 0.035 &   \nd\\
01330349+3042025 & 23.264532 & 30.700701 &  411.09 & 20.863 & 18.972 & 18.044 &    \nd &    \nd & 23.124 & 0.358 & 0.411 &   \nd & 1.056 &  0.200 & -0.014 &   5.92 & 0.005 & 0.039 & 0.030 &   \nd &   \nd & 0.017 &   \nd &   \nd &   \nd\\
01330365+3030497 & 23.265224 & 30.513803 &  395.22 & 20.232 & 18.986 &    \nd &    \nd & 20.722 & 22.154 & 0.365 & 0.248 & 0.494 & 0.219 &  0.145 & -0.051 &    \nd & 0.005 & 0.052 &   \nd &   \nd & 0.005 & 0.005 &   \nd &   \nd &   \nd\\
01330549+3038194 & 23.272890 & 30.638735 &  289.39 & 21.574 & 18.099 & 17.620 &    \nd &    \nd &    \nd & 1.061 & 0.245 &   \nd &   \nd & -0.124 &  0.076 &    \nd & 0.005 & 0.041 & 0.032 &   \nd &   \nd &   \nd &   \nd & 0.050 &   \nd\\
01330596+3037390 & 23.274841 & 30.627512 &  144.20 & 21.362 & 19.027 & 18.253 &    \nd & 22.372 &    \nd & 0.311 & 0.307 & 0.673 &   \nd & -0.148 & -0.279 &    \nd & 0.004 & 0.070 & 0.074 &   \nd & 0.006 &   \nd &   \nd &   \nd &   \nd\\
01330678+3033068 & 23.278263 & 30.551893 &  448.70 & 21.978 & 19.184 &    \nd & 17.483 &    \nd &    \nd & 0.263 & 0.178 &   \nd &   \nd & -0.117 & -0.041 &    \nd & 0.006 & 0.021 &   \nd & 0.047 &   \nd &   \nd &   \nd &   \nd &   \nd\\
01330706+3034548 & 23.279436 & 30.581900 &  297.75 & 22.608 & 18.632 &    \nd & 17.336 &    \nd &    \nd & 2.445 & 0.376 &   \nd &   \nd &  0.206 &  0.017 &    \nd & 0.022 & 0.026 &   \nd & 0.030 &   \nd &   \nd & 0.033 &   \nd &   \nd\\
01330740+3032356 & 23.280830 & 30.543226 &  410.86 & 23.253 &    \nd &    \nd & 18.810 &    \nd &    \nd & 0.520 & 0.118 &   \nd &   \nd & -0.360 & -0.051 &    \nd & 0.019 &   \nd &   \nd & 0.216 &   \nd &   \nd & 0.047 & 0.314 &   \nd\\
01330771+3043034 & 23.282106 & 30.717609 &  188.02 & 21.211 & 19.661 &    \nd & 16.971 & 22.374 &    \nd & 0.352 & 0.248 & 0.493 &   \nd &  0.471 & -0.437 &    \nd & 0.007 & 0.082 &   \nd & 0.054 & 0.005 &   \nd & 0.017 & 0.075 &   \nd\\
01330991+3035253 & 23.291283 & 30.590351 &   95.80 & 22.131 & 19.200 &    \nd & 17.895 &    \nd &    \nd & 0.245 & 0.073 &   \nd &   \nd & -0.151 & -0.128 &    \nd & 0.005 & 0.034 &   \nd & 0.111 &   \nd &   \nd &   \nd &   \nd &   \nd\\
\bottomrule
\end{tabular}

\vspace*{18pt}

\begin{tabular}{lrrrrrrrrrrrrrrrccccccccl}
\toprule
\multicolumn{1}{c}{ID} & \multicolumn{1}{c}{$\sigma_{A_g}$} & \multicolumn{1}{c}{$\sigma_{\phi_i}$} & \multicolumn{1}{c}{$\sigma_{\phi_J}$} & \multicolumn{1}{c}{Type/} & \multicolumn{1}{c}{Class} & \multicolumn{1}{c}{$W_{JK}$} & \multicolumn{1}{c}{$W_{JH}$} & \multicolumn{1}{c}{$\sigma_{W_{JK}}$} & \multicolumn{1}{c}{$\sigma_{W_{JH}}$} & \multicolumn{6}{c}{MLC scores}  & \multicolumn{6}{c}{MLC threshold}& \multicolumn{1}{c}{MT} & \multicolumn{1}{c}{V} & \multicolumn{1}{l}{Fit(s)}\\
& \multicolumn{3}{c}{[mag]} & Rej & & & & & & \multicolumn{1}{c}{LR} & \multicolumn{1}{c}{RAF} & \multicolumn{1}{c}{LDA} & \multicolumn{1}{c}{QDA} & \multicolumn{1}{c}{SVM} & \multicolumn{1}{c}{BSVM}\\
\midrule
01322479+3047175 &    \nd & 0.001 & 0.071 & RH & FA &    \nd &    \nd &   \nd &   \nd &    \nd &    \nd &    \nd &    \nd &    \nd &    \nd &    &    &    &    &    &    & 0 & Z & NA                                             \\
01322533+3036126 &    \nd & 0.007 & 0.045 & LP & FA &    \nd &    \nd &   \nd &   \nd &    \nd &    \nd &    \nd &    \nd &    \nd &    \nd &    &    &    &    &    &    & 0 & Z & NA                                             \\
01322583+3047503 &    \nd & 0.001 & 0.067 & CH & FA &    \nd &    \nd &   \nd &   \nd &    \nd &    \nd &    \nd &    \nd &    \nd &    \nd &    &    &    &    &    &    & 0 & Z & NA                                             \\
01322793+3039553 &    \nd & 0.006 & 0.057 & ML & FO &    \nd &    \nd &   \nd &   \nd &  0.023 &  0.000 &  0.024 &  0.001 &  0.001 &  0.060 &  0 &  0 &  0 &  0 &  0 &  0 & 0 & Z & NA                                             \\
01322879+3049434 &    \nd & 0.001 &   \nd & VZ & FU &    \nd &    \nd &   \nd &   \nd &  0.016 &  0.278 &  0.017 &  0.000 &  0.688 &  0.917 &  0 &  1 &  0 &  0 &  0 &  0 & 1 & P & NA                                             \\
01323222+3041360 &    \nd & 0.002 & 0.009 & NU & FU &    \nd &    \nd &   \nd &   \nd &  0.661 &  0.690 &  0.640 &  0.553 &  0.950 &  0.986 &  1 &  1 &  0 &  0 &  1 &  1 & 4 & Y & HN qHN                                         \\
01323651+3037358 &    \nd & 0.008 & 0.099 & NO & FO &    \nd &    \nd &   \nd &   \nd &  0.566 &  0.278 &  0.827 &  0.838 &  0.831 &  0.968 &  0 &  1 &  1 &  1 &  0 &  1 & 4 & Y & HN qHN                                         \\
01330333+3038314 &    \nd & 0.007 & 0.171 & CJ & FU &    \nd &    \nd &   \nd &   \nd &    \nd &    \nd &    \nd &    \nd &    \nd &    \nd &    &    &    &    &    &    & 0 & Z & NA                                             \\
01330346+3041361 &    \nd & 0.001 & 0.017 & CU & FU &    \nd & 14.688 &   \nd & 0.112 &  0.615 &  0.674 &  0.623 &  0.126 &  0.970 &  0.998 &  1 &  1 &  0 &  0 &  1 &  1 & 4 & Y & JH lJHN lHN lJN qJH qJHN qHN qJ qJN            \\
01330349+3042025 &    \nd & 0.002 & 0.020 & JH & FA &    \nd & 16.442 &   \nd & 0.106 &    \nd &    \nd &    \nd &    \nd &    \nd &    \nd &    &    &    &    &    &    & 0 & Z & NA                                             \\
01330365+3030497 &    \nd & 0.001 & 0.029 & RJ & FA &    \nd &    \nd &   \nd &   \nd &    \nd &    \nd &    \nd &    \nd &    \nd &    \nd &    &    &    &    &    &    & 0 & Z & NA                                             \\
01330549+3038194 &    \nd & 0.001 & 0.015 & RC & FU &    \nd &    \nd &   \nd &   \nd &  1.000 &  1.000 &  1.000 &  1.000 &  1.000 &  1.000 &  1 &  1 &  1 &  1 &  1 &  1 & 6 & M & NA                                             \\
01330596+3037390 &    \nd & 0.001 & 0.040 & OO & FO &    \nd & 16.917 &   \nd & 0.234 &  0.416 &  0.142 &  0.581 &  0.440 &  0.924 &  0.977 &  0 &  0 &  0 &  0 &  1 &  1 & 2 & Y & JH lJHN lH lHN lJ lJN qJH qJHN qHN qJ qJN      \\
01330678+3033068 &    \nd & 0.004 & 0.032 & JK & FA & 16.220 &    \nd & 0.084 &   \nd &    \nd &    \nd &    \nd &    \nd &    \nd &    \nd &    &    &    &    &    &    & 0 & Z & NA                                             \\
01330706+3034548 &    \nd & 0.002 & 0.021 & OU & FU & 16.375 &    \nd & 0.055 &   \nd &  1.000 &  0.740 &  1.000 &  1.000 &  0.832 &  0.952 &  1 &  1 &  1 &  1 &  0 &  1 & 5 & Y & JK lJKN lK lKN lJ lJN qJK qJKN qK qKN qJ qJN   \\
01330740+3032356 &    \nd & 0.004 & 0.481 & RK & FU &    \nd &    \nd &   \nd &   \nd &    \nd &    \nd &    \nd &    \nd &    \nd &    \nd &    &    &    &    &    &    & 0 & Z & NA                                             \\
01330771+3043034 &    \nd & 0.002 &   \nd & CO & FO & 14.975 &    \nd & 0.112 &   \nd &  0.715 &  0.746 &  0.909 &  0.902 &  0.894 &  0.980 &  1 &  1 &  1 &  1 &  1 &  1 & 6 & Y & JK lJKN lK lKN lJ qJK qJKN qK qKN qJ           \\
01330991+3035253 &    \nd & 0.005 & 0.071 & CK & FO &    \nd &    \nd &   \nd &   \nd &    \nd &    \nd &    \nd &    \nd &    \nd &    \nd &    &    &    &    &    &    & 0 & Z & NA\\
\bottomrule
\end{tabular}
}

 \caption{Derived properties and associated uncertainties of all LPVs detected in our analysis. All magnitudes are mean values, based on sinusoidal fits. NIR magnitudes have been corrected for crowding. The full version of this table is available online; only a few representative lines are shown here for guidance. {\bf Type/Rejection code}: CO, Carbon first-overtone; CU, Carbon fundamental-mode; OO, Oxygen first-overtone; OU, Oxygen fundamental-mode; NO, unclassified first-overtone; NU, unclassified fundamental-mode; C[JHK], rejected due to large crowding correction and/or large crowding correction uncertainty in $J$, $H$ or $K_S$; RC, rejected due to abnormally blue colors; R[JHK], rejected as faint outliers in $J$, $H$ or $K_S$ PLRs; JH/JK, rejected as faint outliers in Wesenheit PLRs; LP, rejected due to best-fit period being at limit of grid search; ML, did not pass any classifier thresholds; VZ, rejected by visual inspection. {\bf Class}: FU, fundamental mode; FO, first overtone; FA, faint. {\bf Machine-learning classifier thresholds:} 1, above; 0, below. {\bf MT:} Sum of MLC thresholds: 6, gold sample; $\geq 3$, silver sample, $\geq 1$, bronze sample. {\bf V:} Visual inspection: Y, high quality; M, medium quality; N, low quality; Z, rejected before inspection. {\bf Fit(s):} list of PLRs for which a given variable remained in the final baseline fit after outlier rejection; N/A: not applicable, object rejected before fits.}
 \label{tab:lpvprop}
\end{table*}

\begin{table*}
 \centering
 \begin{tabular}{llrrrrrr}
\toprule
\multicolumn{1}{c}{Host}  & \multicolumn{1}{c}{Mag} &
\multicolumn{1}{c}{$a_0$} & \multicolumn{1}{c}{$\sigma(a_0)$} & \multicolumn{1}{c}{$a_1$} & \multicolumn{1}{c}{$\sigma(a_1)$} &
\multicolumn{1}{c}{$N$}   & \multicolumn{1}{c}{$\sigma$}      \\
 &   &  \multicolumn{2}{c}{[mag]} & \multicolumn{2}{c}{[mag/dex]} & & \multicolumn{1}{c}{[mag]}\\

\midrule
LMC & $W_{JK_S}$ & 10.743 & 0.020 & -3.649 & 0.157 &  163 & 0.227 \\
LMC & $W_{JH}$   & 10.696 & 0.029 & -3.347 & 0.216 &  163 & 0.316 \\
LMC & $K_S$      & 11.560 & 0.019 & -3.368 & 0.152 &  163 & 0.220 \\
LMC & $H$        & 11.942 & 0.023 & -3.161 & 0.175 &  163 & 0.254 \\
LMC & $J$        & 12.640 & 0.022 & -3.142 & 0.162 &  163 & 0.237 \\
\midrule
M33 & $W_{JK_S}$ & 16.826 & 0.007 & \multicolumn{2}{c}{\nd} & 578 & 0.166 \\
M33 & $W_{JH}$   & 16.660 & 0.013 & \multicolumn{2}{c}{\nd} & 534 & 0.283 \\
M33 & $K_S$      & 17.676 & 0.006 & \multicolumn{2}{c}{\nd} & 593 & 0.140 \\
M33 & $H$        & 18.029 & 0.006 & \multicolumn{2}{c}{\nd} & 537 & 0.151 \\
M33 & $J$        & 18.827 & 0.006 & \multicolumn{2}{c}{\nd} & 781 & 0.173 \\
\bottomrule
\end{tabular}
 \caption{Coefficients of the linear P-L relations fit to O-rich fundamental-mode Miras with $P\!<\!400$~d in the LMC{\ that satisfied the color-color relations from Table~\ref{tab:colcol}}, and intercepts for the corresponding M33 samples when holding the slopes fixed to the LMC values.}
 \label{tab:ogle_linear_plr_coefficients}
\end{table*}

\begin{table*}
 \centering
 \begin{tabular}{lrrrrrr}
\toprule
\multicolumn{1}{c}{Mag} & \multicolumn{1}{c}{$a_0$} & \multicolumn{1}{c}{$\sigma(a_0)$} & \multicolumn{1}{c}{$a_1$} & \multicolumn{1}{c}{$\sigma(a_1)$} & \multicolumn{1}{c}{$N$}   & \multicolumn{1}{c}{$\sigma$}      \\
  &  \multicolumn{2}{c}{[mag]} & \multicolumn{2}{c}{[mag/dex]} & & \multicolumn{1}{c}{[mag]}\\
\midrule
$W_{JK_S}$& 16.981 & 0.008 & -4.434 & 0.054 & 790 & 0.177 \\
$W_{JH}$  & 16.886 & 0.015 & -4.174 & 0.103 & 806 & 0.231 \\  
$K_S$     & 17.764 & 0.004 & -3.700 & 0.026 & 816 & 0.184 \\   
$H$       & 18.142 & 0.004 & -3.334 & 0.027 & 821 & 0.199 \\   
$J$       & 18.880 & 0.004 & -3.060 & 0.028 & 798 & 0.209 \\   
\midrule
$W_{JK_S}$ & 16.882 & 0.007 & \multicolumn{2}{c}{\nd} & 797 & 0.195 \\
$W_{JH}$   & 16.785 & 0.009 & \multicolumn{2}{c}{\nd} & 801 & 0.238 \\
$K_S$      & 17.725 & 0.006 & \multicolumn{2}{c}{\nd} & 821 & 0.192 \\
$H$        & 18.125 & 0.006 & \multicolumn{2}{c}{\nd} & 816 & 0.197 \\
$J$        & 18.889 & 0.007 & \multicolumn{2}{c}{\nd} & 798 & 0.208 \\
\bottomrule
\end{tabular}
 \caption{PLRs for O-rich, fundamental-mode Miras in M33 with $125<P<400$~d based on magnitudes from \citet{yuan2018} and solving for a slope (top set) or holding the slope fixed to the LMC values from Table~\ref{tab:ogle_linear_plr_coefficients} (bottom set).}
 \label{tab:plry18}
\end{table*}

\begin{table*}
 \begin{tabular}{lllrrrrrr}
\toprule
Band & Sample & Slope & \multicolumn{1}{c}{$a_0$} & \multicolumn{1}{c}{$\sigma(a_0)$} & \multicolumn{1}{c}{$a_1$} & \multicolumn{1}{c}{$\sigma(a_1)$} & \multicolumn{1}{c}{$N$} & \multicolumn{1}{c}{$\sigma$} \\
     &        &       & \multicolumn{2}{c}{[mag]} & \multicolumn{2}{c}{[mag/dex]} & & \multicolumn{1}{c}{[mag]}\\
\midrule
\multirow{4}{3em}{$g$} & \multirow{2}{4.5em}{M33-ML}   & Free  & 23.045 & 0.058 & 0.829 & 0.714 &  91 & 0.584 \\
                       &                               & Fixed & 23.033 & 0.063 & 3.830 & 2.378 &  92 & 0.645 \\
                       &  \multirow{2}{4.5em}{M33-LMC} & Free  & 23.049 & 0.058 & 0.857 & 0.702 &  91 & 0.585 \\
                       &                               & Fixed & 23.037 & 0.063 & 3.830 & 2.378 &  92 & 0.644 \\
\midrule         
\multirow{4}{3em}{$r$} &  \multirow{2}{4.5em}{M33-ML}  & Free  & 22.289 & 0.031 & 2.520 & 0.354 & 286 & 0.523 \\
                       &                               & Fixed & 22.296 & 0.030 & 1.892 & 1.816 & 300 & 0.518 \\
                       & \multirow{2}{4.5em}{M33-LMC}  & Free  & 22.286 & 0.031 & 2.311 & 0.353 & 283 & 0.531 \\
                       &                               & Fixed & 22.290 & 0.030 & 1.892 & 1.816 & 280 & 0.524 \\
\midrule         
\multirow{4}{3em}{$i$} & \multirow{2}{4.5em}{M33-ML}   & Free  & 21.109 & 0.016 & 1.936 & 0.142 & 531 & 0.327 \\
                       &                               & Fixed & 21.201 & 0.015 & 0.653 & 1.462 & 552 & 0.382 \\
                       & \multirow{2}{4.5em}{M33-LMC}  & Free  & 21.113 & 0.016 & 1.928 & 0.141 & 528 & 0.324 \\
                       &                               & Fixed & 21.201 & 0.015 & 0.653 & 1.462 & 542 & 0.369 \\
\bottomrule
\end{tabular}

 \caption{Coefficients associated with the linear $gri$ PLRs fit to the unique, O-rich candidates with $P<400$~d identified using machine learning classifiers (the ML-M33 sample) and the O-rich candidates with $P<400$~d identified by fitting LMC-based PLRs to the M33 Mira candidates (the LMC-M33 sample). The ``Slope'' columns indicates whether $a_1$ was kept fixed or allowed to vary. The $a_1$ values and uncertainties for the ``Fixed'' rows are from \citet{iwanek2021}.}
 \label{tab:gri_params}
\end{table*}

\begin{figure*}
 \includegraphics[height=0.35\textheight]{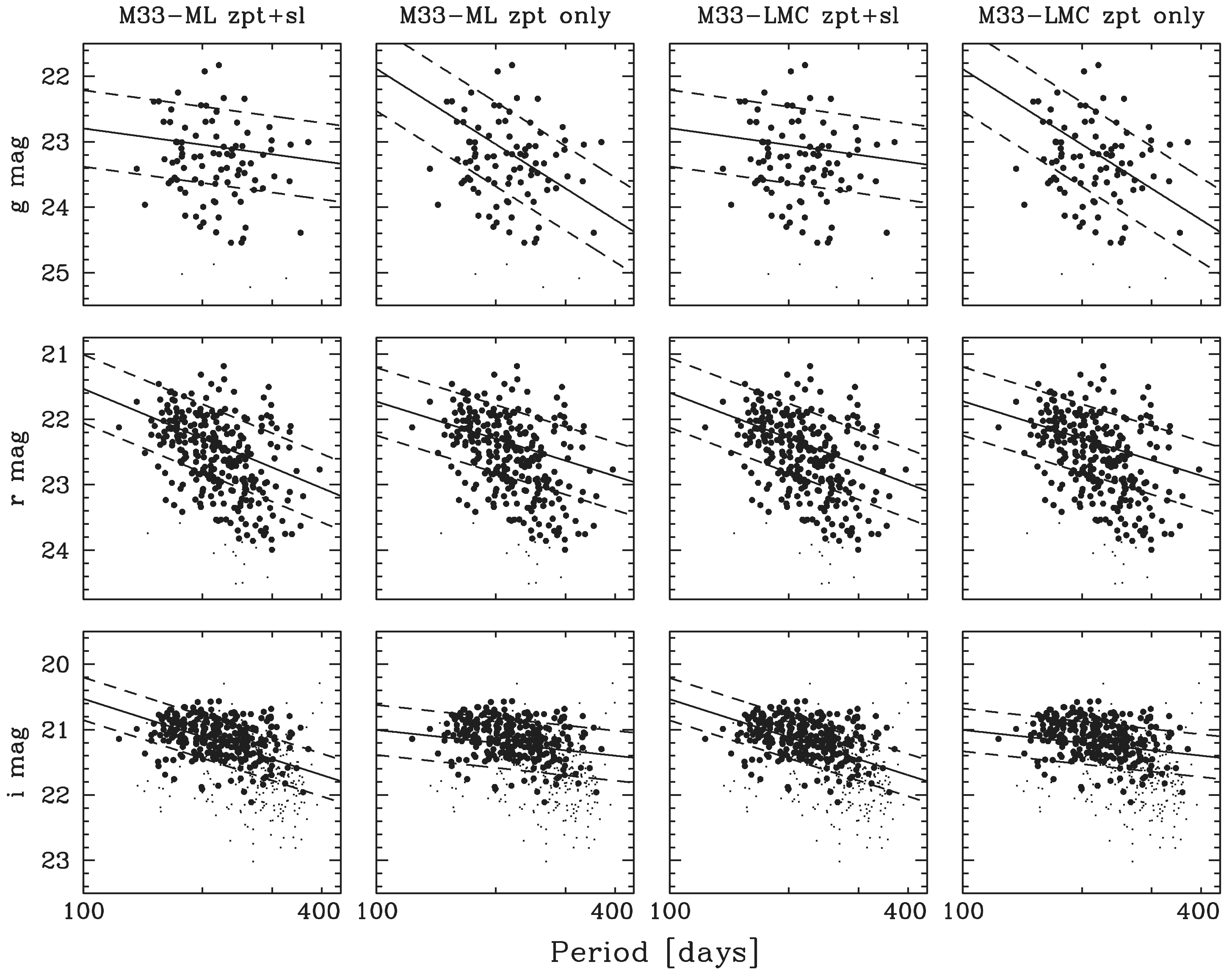}
 \caption{Linear PLRs in $g$ (top), $r$ (middle) and $i$ (bottom) for fundamental-mode O-rich Mira candidates with $P<400$~d identified in \S \ref{sec:ml_classifiers} (M33-ML; left two columns) and \S\ref{sec:ogle_m33_plr} (M33-LMC; right two columns). For each sample, the left column shows a fit for zeropoint and slope, while the right one fixes the slopes to the values determined by \citet{iwanek2021}.}
 \label{fig:gri_plrs}
\end{figure*}
\end{document}